%% file: MSTM_maindoc_and_supp.tex
\documentclass[12pt]{article}
\usepackage{setspace}
\usepackage{float}
\usepackage{amssymb}
\usepackage{amsthm}
\usepackage{amsmath}
\usepackage{latexsym}
\usepackage{epsfig}
\usepackage{graphicx}
\usepackage{wasysym}
\usepackage{threeparttable}
\usepackage{natbib}
\usepackage{color}
\usepackage{epstopdf}
\usepackage{caption}
\usepackage{subcaption}

\usepackage[breaklinks]{hyperref}

\newcommand{\ind}{\stackrel{\mathrm{ind}}{\sim}}

\def\boxit#1{\vbox{\hrule\hbox{\vrule\kern6pt
          \vbox{\kern6pt#1\kern6pt}\kern6pt\vrule}\hrule}}

\def\bse{\begin{eqnarray*}}
\def\ese{\end{eqnarray*}}
\def\be{\begin{eqnarray}}
\def\ee{\end{eqnarray}}
\def\bq{\begin{equation}}
\def\eq{\end{equation}}
\def\bse{\begin{eqnarray*}}
\def\ese{\end{eqnarray*}}

\usepackage{mathptmx}      
\usepackage{bm}

 {\begin{list}{}%
         {\setlength{\leftmargin}{#1}}%
         \item[]%
 }
 {\end{list}}

\include{notations4}

\begin{document}
\thispagestyle{empty} \baselineskip=28pt

\begin{center}
{\LARGE{\bf Mixed Effects Modeling for Areal Data that Exhibit Multivariate-Spatio-Temporal Dependencies}}
\end{center}

\baselineskip=12pt

\vskip 2mm
\begin{center}
Jonathan R. Bradley\footnote{(\baselineskip=10pt to whom correspondence should be addressed) Department of Statistics, University of Missouri, 146 Middlebush Hall, Columbia, MO 65211, bradleyjr@missouri.edu},
Scott H. Holan\footnote{\baselineskip=10pt  Department of Statistics, University of Missouri, 146 Middlebush Hall, Columbia, MO 65211-6100},
Christopher K. Wikle$^2$
\end{center}
%
%
%
%
\vskip 4mm

\begin{center}
\large{{\bf Abstract}}
\end{center}
There are many data sources available that report related variables of interest that are {also} referenced over geographic regions and time; however, there are relatively few general statistical methods that one can readily use that incorporate these multivariate-spatio-temporal dependencies. As such, we introduce the {multivariate-spatio-temporal} mixed effects model (MSTM) to analyze areal data with multivariate-spatio-temporal dependencies. The proposed MSTM extends the notion of Moran's I basis functions to the multivariate-spatio-temporal setting. This extension leads to several methodological contributions including extremely effective dimension reduction, a dynamic linear model for multivariate-spatio-temporal areal processes, and the reduction of a high-dimensional parameter space using {a novel} parameter model. Several examples are used to demonstrate that the MSTM provides an extremely viable solution to many important problems found in different and distinct corners of the spatio-temporal statistics literature including: modeling nonseparable and nonstationary covariances, combing data from multiple repeated surveys, and analyzing massive multivariate-spatio-temporal datasets.
\baselineskip=12pt

%
%
%

\baselineskip=12pt
\par\vfill\noindent
{\bf Keywords:} { American Community Survey;} {Longitudinal Employer-Household Dynamics (LEHD) program;} Kalman filter; Markov chain Monte Carlo; Multivariate spatio-temporal data; Moran's I basis.
\par\medskip\noindent
\clearpage\pagebreak\newpage \pagenumbering{arabic}
\baselineskip=24pt

\section{Introduction}\label{sec:intro}

Ongoing data collection from the private sector along with federal, state, and local governments have produced massive quantities of data measured over geographic regions (areal data) and time. This unprecedented volume of spatio-temporal data contains a wide range of variables and has, thus, created unique challenges and opportunities for those practitioners seeking to capitalize on their full utility. For example, methodological challenges arise because these data exhibit variability over multiple processes (latent variables), geographies, and time-points. As such, the corresponding multivariate-spatio-temporal covariances may be quite complex, involving nonstationarity and interactions between different variables, regions, and times. To effectively model these complex dependencies we introduce the multivariate-spatio-temporal mixed effects model (MSTM).

 Despite the wide availability of areal datasets exhibiting multivariate-spatio-temporal dependencies, the literature on modeling multivariate-spatio-temporal areal processes is relatively recent by comparison. For example, versions of a multivariate space-time conditional autoregressive (CAR) model have been proposed by \citet{carlinmst}, \citet{congdon}, \citet{pettitt}, \citet{zhuglm}, \citet{daniels}, and \citet{bestmst}, among others. However, these methodologies cannot accommodate data from multiple repeated surveys, and cannot efficiently model very-large-to-massive datasets. Additionally, these approaches impose separability assumptions (and in some cases independence), which are not appropriate for many settings, as these models fail to capture important interactions (and dependence) between different variables, regions, and times \citep{steinSep}.
 
 In addressing the aforementioned issues, the proposed MSTM provides several methodological contributions, including a novel class of multivariate-spatio-temporal basis functions. Additionally, the MSTM framework introduces an extremely powerful dimension reduction approach for the multivariate-spatio-temporal setting. Specifically, we introduce a novel conditional (i.e., first-order) multivariate-spatio-temporal-dynamic-linear model specification, and propose an innovative parameter model to reduce a high-dimensional parameter space.

The foundation for these methodological advances consists of two general techniques for modeling dependencies. The first technique allows different random processes (latent variables) to share the same random effect. This idea has been effectively utilized in the spatial \citep{banerjee,johan,lindgren-2011,hughes,nychkaLK}, multivariate-spatial \citep{royle1999,finley,finley2}, and spatio-temporal \citep{stcar,wikle2001} settings.   The second technique we consider partitions the joint likelihood into a product of more manageable conditional likelihoods; see, \citet{berlinerroyle} for the multivariate spatial setting, and \citet[][Chap. 7]{cressie-wikle-book} for the spatio-temporal setting. 
  
To incorporate multivariate spatial dependencies in our MSTM framework we let the $L$ different variables under consideration share the same random effects at time $t$ (i.e., the first technique). Then, to build in temporal dependence we use a dynamic Bayesian hierarchical model (BHM) and specify conditional distributions for the temporal random effects at time $t$ given previous times (i.e., the second technique). Together, these two techniques are combined in a novel way to define the MSTM. In what follows, we demonstrate that our approach is extremely general and that many datasets can be effectively modeled using the MSTM framework.

Accounting for the possibility of ``big data'' is arguably one of the most important features to consider, since the size of modern spatio-temporal datasets creates unavoidable methodological difficulties. In fact, there is a large literature available on modeling large-to-massive spatial/spatio-temporal datasets and similar complications arise in the multivariate-spatio-temporal, though often exacerbated. The primary issue surrounding this literature is that there is a computational bottleneck that occurs when computing a {high-dimensional} Gaussian likelihood. {See \citet{cressie-wikle-book}, \citet{reviewmethods}, and the references therein for a comprehensive review of spatial and spatio-temporal models for ``big data.''} To illustrate the exceptional utility of our approach, we demonstrate that the MSTM can efficiently model massive data by considering a survey dataset consisting of 7,530,037 observations and 3,680 spatial fields, which we jointly model using the MSTM.

We are able to analyze this massive dataset using the MSTM framework, since the model can be specified to have a computationally advantageous reduced rank structure \citep[e.g., see][]{wikleHandbook}. This reduced rank structure is achieved, in part, by extending aspects of the model suggested by \citet{hughes} from the univariate spatial-only setting to the multivariate-spatio-temporal setting. Specifically, we extend the Moran's I (MI) basis functions to the multivariate-spatio-temporal setting \citep[see][for the spatial only case]{griffith2000, griffith2002, griffith2004, griffith2007,hughes,aaronp}. Further, we introduce a novel propagator matrix for a first-order vector autoregressive (VAR(1)) model, which we call the MI propagator matrix. Here, the propagator matrix of {the} VAR(1) is specified to have a desirable non-confounding property, which is similar to the specification of the MI basis functions in \citet{hughes}.

In addition to the MI basis functions and propagator matrix, we also propose an extension of the spatial random effects covariance parameter model used in \citet{hughes} and \citet{aaronpBayes}, which we call the MI prior. Here, we interpret the MI prior as a rescaling of the covariance matrix that is specified to be close (in Frobenius norm) to a ``target precision'' matrix. This parameterization significantly reduces the dimension of the parameter space, thereby reducing the computation needed for fully Bayesian inference. Furthermore, this target precision matrix can be sensibly chosen based on knowledge of the underlying spatial process.

 Given these methodological advances, the MSTM can be used to effectively address numerous statistical modeling and analysis problems in the context of multivariate-spatio-temporal areal data.  In particular, we use the MSTM to model nonseparable and nonstationary covariances, to combine data from multiple repeated surveys, and to analyze a massive multivariate-spatio-temporal areal dataset.  Although, in this article, we mainly focus on these particular applications, the MSTM is tremendously flexible and can be readily adapted to other settings.

The remainder of this article is organized as follows. In Section~\ref{sec:MSTM}, we introduce and define the MSTM, which includes our proposed MI propagator matrix. Then, in Section~\ref{sec:parmod} we introduce the parameter model for the covariance matrix of the random effects term and show explicitly how one can incorporate knowledge of the spatial process into this parameter model. Next, in Section~\ref{sec:app} we demonstrate the use of the MSTM through three distinct modeling applications,  namely: analyzing cancer mortality rates using data obtained from the National Cancer Institute (NCI); combining unemployment rates from the American Community Survey (ACS, US Census Bureau) and the Local Area Unemployment Survey  (LAUS, Bureau of Labor Statistics (BLS)), and modeling multivariate-spatio-temporal data from the Longitudinal Employer-Household Dynamics (LEHD) program (US Census Bureau). These examples are used to demonstrate that we can model nonseparable and nonstationary multivariate-spatio-temporal covariances, combine data from multiple surveys, and process massive data using the MSTM. We end with a discussion in Section~\ref{sec:disc}. For convenience of exposition, details surrounding some of the technical results and the MCMC algorithm are left to the Appendix.

\section{The Multivariate-Spatio-Temporal Mixed Effects Model}\label{sec:MSTM}
The MSTM is defined hierarchically using the familiar data model, process model, and parameter model organization \citep{berlinhier,cressie-wikle-book}. In this section, we provide the specific details.

\subsection{The Data Model} The data model for the MSTM is defined as,
\begin{equation}
\label{data:model}
Z_{t}^{(\ell)}(A) = Y_{t}^{(\ell)}(A) + \epsilon_{t}^{(\ell)}(A);\hspace{5pt} \ell = 1,\ldots,L, \hspace{5pt}t = T_{L}^{(\ell)},\ldots,T_{U}^{(\ell)}, \hspace{5pt}A \in D_{\mathrm{P},t}^{(\ell)},
\end{equation}
\noindent
where $\{Z_{t}^{(\ell)}(\cdot)\}$ represents multivariate-spatio-temporal data. The components of (\ref{data:model}) are defined and elaborated as follows:
\begin{enumerate}
	\item The subscript ``$t$'' denotes discrete time, and the superscript ``$\ell$'' indexes different variables of interest (e.g., unemployment rates). There are a total of $L$ variables of interest (i.e., $\ell = 1,\ldots,L$) and we allow for a different number of observed time-points for each of the $L$ variables of interest (i.e., for variable $\ell$, $t = T_{L}^{(\ell)},\ldots,T_{U}^{(\ell)}$).
	\item We require $T_{L}^{(\ell)},\ldots,T_{U}^{(\ell)}$ to be on the same temporal scale (e.g., monthly, quarterly, etc.) for each $\ell$, $T_{L}^{(\ell)} \le T_{U}^{(\ell)}$, $\mathrm{min}\left(T_{L}^{(\ell)}\right) = 1$, and $\mathrm{max}\left(T_{U}^{(\ell)}\right) = T \ge 1$. 
	\item The set $A$ represents a generic areal unit. For example, a given set $A$ might represent a state, county, or a Census tract. Denote the collection of all $n_{t}^{(\ell)}$ observed areal units with the set $D_{\mathrm{O},t}^{(\ell)}\equiv \{A_{t,i}^{(\ell)}: i = 1,\ldots,n_{t}^{(\ell)}\}$; $\ell = 1,\ldots,L$. The observed data locations are different from the prediction locations $D_{\mathrm{P},t}^{(\ell)}\equiv \{A_{t,j}^{(\ell)}: j = 1,\ldots,N_{t}^{(\ell)}\}$; that is, we consider predicting on a spatial support that may be different from $\{D_{\mathrm{O},t}^{(\ell)}\}$. Additionally, denote the number of prediction locations at time $t$ as $N_{t} =\sum_{\ell = 1}^{L}N_{t}^{(\ell)}$, and the total number of prediction locations as ($n \equiv \sum_{t = 1}^{T}n_{t}$) $N \equiv \sum_{t = 1}^{T}N_{t}$. In a similar manner, the number of observed locations at time $t$ and total number of observations are given by $n_{t} =\sum_{\ell = 1}^{L}n_{t}^{(\ell)}$ and $n \equiv \sum_{t = 1}^{T}n_{t}$, respectively.
	\item The random process $Y_{t}^{(\ell)}(\cdot)$ represents the $\ell$-th variable of interest at time $t$. For example, $Y_{t}^{(\ell)}(\cdot)$ might represent the cancer mortality rate for females at time $t$. The stochastic properties of $\{Y_{t}^{(\ell)}(\cdot)\}$ are defined in Section~\ref{sec:process}. Latent processes like $\{Y_{t}^{(\ell)}(\cdot)\}$ have been used to incorporate spatio-temporal dependencies \citep[e.g., see][]{cressie-wikle-book}, which we modify to the multivariate-spatio-temporal setting.
	\item It is assumed that $\epsilon_{t}^{(\ell)}(\cdot)$ is a white-noise Gaussian process with mean zero and known variance var(\(\epsilon_{t}^{(\ell)}(\cdot)\))= $v_{t}^{(\ell)}(\cdot)$ for $\ell = 1,\ldots,L,$ and $t = T_{L}^{(\ell)},\ldots,T_{U}^{(\ell)}$. The presence of $\{\epsilon_{t}^{(\ell)}(\cdot)\}$ in (\ref{data:model}) allows us to take into account that we do not perfectly observe $\{Y_{t}^{(\ell)}(\cdot)\}$, and instead observe a {noisy} version $\{Z_{t}^{(\ell)}(\cdot)\}$. In many settings, there is information that we can use to define $\{\epsilon_{t}^{(\ell)}(\cdot)\}$ (e.g., information provided by a statistical agency). In particular, {variances} are provided by the statistical agency {(e.g., NCI, US Census Bureau, and BLS provide survey variance estimates)}. If one does not account for this extra source of variability
	then the total variability of the process $\{Y_{t}^{(\ell)}(\cdot)\}$ may be underestimated. For example, \citet{finley} show that if one ignores white-noise error in a Gaussian linear model then one underestimates the total variability of the latent process of interest.
\end{enumerate}

\subsection{The Process Model}\label{sec:process}
The process model for MSTM is defined as,
\begin{align}
\label{process:model}
Y_{t}^{(\ell)}(A) = \mu_{t}^{(\ell)}(A) + \textbf{S}_{t}^{(\ell)}(A)^{\prime}\bm{\eta}_{t} + \xi_{t}^{(\ell)}(A);&\hspace{5pt} \ell = 1,\ldots,L, t = T_{L}^{(\ell)},\ldots,T_{U}^{(\ell)},A \in D_{\mathrm{P},t}^{(\ell)}.
\end{align}
\noindent
In (\ref{process:model}), $Y_{t}^{(\ell)}(\cdot)$ represents the $\ell$-th spatial random process of interest at time $t$, which is modeled by three terms on the right-hand side of (\ref{process:model}). The first term (i.e., $\{\mu_{t}^{(\ell)}(\cdot)\}$) is a fixed effect, which is unknown, and requires estimation. We set $\mu_{t}^{(\ell)}(\cdot)\equiv \textbf{x}_{t}^{(\ell)}(\cdot)^{\prime}\bm{\beta}_{t}$, where $\textbf{x}_{t}^{(\ell)}$ is a known $p$-dimensional vector of covariates and $\bm{\beta}_{t}\in \mathbb{R}^{p}$ is a fixed unknown parameter vector; $\ell = 1,\dots,L$ and $t = 1,\ldots,T$. In general, we allow both $\textbf{x}_{t}^{(\ell)}$ and $\bm{\beta}_{t}$ to change over time; however, in practice one must assess whether or not this is appropriate for their application.

The second term on the right-hand side of (\ref{process:model}) (i.e., $\{\textbf{S}_{t}^{(\ell)}(\cdot)^{\prime}\bm{\eta}_{t}\}$) represents multivariate-spatio-temporal dependencies; in Section~\ref{sec:dynamics}, we provide the stochastic properties of $\{\bm{\eta}_{t}\}$. Here, the $r$-dimensional vectors of multivariate-spatio-temporal basis functions $\textbf{S}_{t}^{(\ell)}(\cdot)\equiv (S_{t,1}^{(\ell)}(\cdot),\ldots,S_{t,r}^{(\ell)}(\cdot))^{\prime}$ are pre-specified for each $t = 1,\ldots,T$ and $\ell = 1,\ldots,L$. {In principle,} the $r$-dimensional vector $\textbf{S}_{t}^{(\ell)}(\cdot)$ can belong to any class of spatial basis functions; however, we shall use the Moran's I (MI) basis functions \citep{griffith2000, griffith2002, griffith2004, griffith2007,hughes,aaronp}. The MI basis functions are a class of functions used to model areal data in a reduced dimensional space (i.e., $r \ll n$). This feature allows for fast computation of the distribution of $\{\bm{\eta}_{t}\}$, which can become computationally expensive for large $r$ \citep{hughes, aaronp}. This will be particularly useful for the datasets in Section~\ref{sec:app}, which {for one of our examples has 7,530,037 observations}. Additionally, the MI basis functions allow for nonstationarity in space and, for areal data, this is a desirable property (e.g., see \citet{banerjee-etal-2004} and the references therein).

Now, the $r$-dimensional vectors of MI basis functions $\{\textbf{S}_{t}^{(\ell)}(\cdot)\}$ are equivalent to the first $r$ eigenvectors of the MI operator \citep[see,][]{hughes}. That is, the MI operator at time $t$ is defined as
\begin{equation}
\textbf{G}(\textbf{X}_{t},\textbf{A}_{t}) \equiv\left(\textbf{I}_{N_{t}} - \textbf{X}_{t}\left(\textbf{X}_{t}^{\prime}\textbf{X}_{t}\right)^{-1}\textbf{X}_{t}^{\prime}\right)\textbf{A}_{t}\left(\textbf{I}_{N_{t}} - \textbf{X}_{t}\left(\textbf{X}_{t}^{\prime}\textbf{X}_{t}\right)^{-1}\textbf{X}_{t}^{\prime}\right);\hspace{5pt}t = 1,\ldots,T,
\end{equation}
where the $N_{t}\times p$ matrix $\textbf{X}_{t} \equiv \left(\textbf{x}_{t}^{(\ell)}(A): \ell = 1,\ldots,L, A \in D_{\mathrm{P},t}^{(\ell)}\right)^{\prime}$, $\textbf{I}_{N_{t}}$ is an $N_{t}\times N_{t}$ identity matrix, and $\textbf{A}_{t}$ is the $N_{t}\times N_{t}$ adjacency matrix corresponding to the edges formed by $\{D_{\mathrm{P},t}^{(\ell)}:\ell = 1,\ldots,L\}$. From the spectral representation, $\textbf{G}(\textbf{X}_{t},\textbf{A}_{t}) = \bm{\Phi}_{X,G,t}\bm{\Lambda}_{X,G,t}\bm{\Phi}_{G,t}^{\prime}$, we denote the $N_{t}\times r$ real matrix formed from the  first $r$ columns of $\bm{\Phi}_{X,G,t}$ as $\textbf{S}_{X,t}$. Additionally, set the row of $\textbf{S}_{X,t}$ that corresponds to variable $\ell$ and areal unit $A$ equal to $\textbf{S}_{t}^{(\ell)}(A)$. 

The third term on the right-hand side of (\ref{process:model}) (i.e., $\{\xi_{t}^{(\ell)}(\cdot)\}$) represents fine-scale variability and is assumed to be Gaussian white-noise with mean-zero and unknown variance $\{\sigma_{\xi,t}^2\}$. In general, $\{\xi_{t}^{(\ell)}(\cdot)\}$ represents the left-over variability not accounted for by $\{\textbf{S}_{t}^{(\ell)}(\cdot)^{\prime}\bm{\eta}_{t}\}$. In the setting where the variance $\{v_{t}^{(\ell)}(\cdot)\}$ is unknown, one should model the sum $\{\epsilon_{t}^{(\ell)}(\cdot) + \xi_{t}^{(\ell)}(\cdot)\}$ due to identifiability issues between $\{\epsilon_{t}^{(\ell)}(\cdot)\}$ and $\{\xi_{t}^{(\ell)}(\cdot)\}$ \citep[e.g., see][]{banerjee, finley,finley2}. Minor adjustments can be made to our methodology to allow for this. One might also consider modeling spatial covariances in $\{\xi_{t}^{(\ell)}(\cdot)\}$. Again, minor adjustments to our methodology could be used to incorporate, for example, a CAR model \citep[][Chap. 3]{banerjee-etal-2004}, tapered covariances \citep[][pg. 108]{cressie} or block diagonal covariances \citep{steinr} in $\{\xi_{t}^{(\ell)}(\cdot)\}$.
 
\subsection{Temporal Dynamics for the Latent Process}\label{sec:dynamics} We assume $\bm{\eta}_{t}$ is generated using a VAR(1) {model} \citep[][Chap. 7]{cressie-wikle-book}:
\begin{equation}\label{autoregressive}
\bm{\eta}_{t} = \textbf{M}_{t}\bm{\eta}_{t-1} + \bu_{t};\hspace{5pt}t = 2,3,\ldots,T
\end{equation}
\noindent
where for $t = 1,2,\ldots,T$ the $r$-dimensional random vector $\bm{\eta}_{t}$ is Gaussian with mean-zero and has an unknown $r\times r$ covariance matrix $\textbf{K}_{t}$; $\textbf{M}_{t}$ is a $r \times r$ known propagator matrix (see discussion below); and $\bu_{t}$ is an $r$-dimensional Gaussian random vector with mean-zero and unknown $r\times r$ covariance matrix $\textbf{W}_{t}$, and is independent of $\bm{\eta}_{t-1}$.

 First order vector autoregressive models may offer more realistic structure with regards to interactions across space and time. This is a feature that cannot be included in the alternative modeling approaches discussed in Section~\ref{sec:intro}. Additionally, the VAR(1) model has been shown to perform well (empirically) in terms of both estimation and prediction for surveys repeated over time \citep{Jones,Bell,Feder}.

The $r$-dimensional random vectors $\{\bm{\eta}_{t}\}$ are not only used to model temporal dependencies in $\{Y_{t}^{(\ell)}(\cdot)\}$, but are also used to model multivariate dependencies. Notice that the random effect term $\bm{\eta}_{t}$ is common across all $L$ processes. Allowing for a common random effect term between different processes is a simple way to induce dependence \citep[][Chap. 7.4]{cressie-wikle-book}. This strategy has been used by \citet{royle1999}, \citet{finley}, and \citet{finley2} in the multivariate spatial setting { and has been extended here.}

We are now left to specify the $r\times r$ real matrices in the set $\{\textbf{M}_{t}\}$. The problem of confounding provides motivation for the definition of the MI basis functions $\{\textbf{S}_{X,t}^{(\ell)}(\cdot)\}$ \citep{griffith2000, griffith2002, griffith2004, griffith2007,Reich,hughes}. In a similar manner, the problem of confounding manifests in a spatio-temporal VAR(1) model and can be addressed through careful specification of $\{\textbf{M}_{t}\}$. To see this, substitute (\ref{autoregressive}) into (\ref{process:model}) to obtain,
\begin{equation}\label{matrix:process}
\by_{t} = \textbf{X}_{t}\bm{\beta}_{t} + \textbf{S}_{X,t}\textbf{M}_{t}\bm{\eta}_{t-1} + \textbf{S}_{X,t}\bu_{t} + \bm{\xi}_{t};\hspace{5pt} t = 2,\ldots, T
\end{equation}
where the $N_{t}$-dimensional latent random vectors $\by_{t}\equiv (Y_{t}^{(\ell)}(A): \ell = 1,\ldots,L, A \in D_{\mathrm{P},t}^{(\ell)})^{\prime}$ and $\bm{\xi}_{t}\equiv (\xi_{t}^{(\ell)}(A): \ell = 1,\ldots,L, A \in D_{\mathrm{P},t}^{(\ell)})^{\prime}$. The specification of $\{\textbf{S}_{X,t}\}$ using MI basis functions implies that there are no issues with confounding between $\{\bm{\beta}_{t}\}$ and $\{\bu_{t}\}$; however, depending on our choice for $\{\textbf{M}_{t}\}$ there might be issues with confounding between $\bm{\eta}_{t-1}$ and the $(p+r)$-dimensional random vector $\bm{\zeta}_{t} \equiv (\bm{\beta}_{t}^{\prime}, \bu_{t}^{\prime})^{\prime}$; $t = 2,\ldots,T$. Then, rewriting (\ref{matrix:process}), we get
\begin{equation}\label{matrix:process2}
\textbf{S}_{X,t}^{\prime}(\by_{t}-\bm{\xi}_{t}) = \textbf{B}_{t}\bm{\zeta}_{t} + \textbf{M}_{t}\bm{\eta}_{t-1};\hspace{5pt} t = 2,\ldots, T,
\end{equation}
where the $r\times(p+r)$ matrix $\textbf{B}_{t}\equiv (\textbf{S}_{X,t}^{\prime}\textbf{X}_{t}, \textbf{I})$. The representation in (\ref{matrix:process2}) gives rise to what we call the MI propagator matrix, which is defined in an analogous manner to the MI basis functions. Using the spectral representation  of $\textbf{G}(\textbf{B}_{t},\textbf{I}_{r}) = \bm{\Phi}_{G,B,t}\bm{\Lambda}_{G,B,t}\bm{\Phi}_{G,B,t}^{\prime}$ we set the $r\times r$ real matrix $\textbf{M}_{t}$ equal to the first $r$ columns of $\bm{\Phi}_{G,B,t}$ for each $t$, which is denoted with $\textbf{M}_{B,t}$. 

Notice that there are no restrictions on $\{\textbf{M}_{B,t}\}$ to {mathematically guarantee} that $\textbf{M}_{B,t}$ does not become ``explosive'' as $t$ increases. Thus, one should investigate whether or not this is the case when using this model for forecasting. One should also be aware that the covariates $\{\textbf{X}_{t}\}$ inform the MI propagator matrices. Importantly, $\{\textbf{M}_{B,t}\}$'s (functional) dependence with $\{\textbf{X}_{t}\}$ and $\{\textbf{S}_{X,t}\}$ implies nonstationarity in time.

Also, notice that we do not treat $\textbf{M}_{t}$ as an unknown parameter matrix to be estimated. Instead, we chose a specific form for $\{\textbf{M}_{t}\}$, namely $\{\textbf{M}_{B,t}\}$, that avoids confounding between $\{\bm{\eta}_{t}\}$ and $\{\bm{\zeta}_{t}\}$. As a result, the final form of $\{\textbf{M}_{B,t}\}$ might not be interpretable from a spatial point of view. We address this in Section~\ref{sec:parmod}, where constraints are added to the parameter model so that $\mathrm{cov}(\bm{\eta}_{t}) = \textbf{M}_{B,t}\textbf{K}_{t-1}\textbf{M}_{B,t}^{\prime} + \textbf{W}_{t}$ is { spatially interpretable.}

\section{The Parameter Model}\label{sec:parmod} 
At this stage, one could specify any desired prior for the $r\times r$ covariance matrices $\textbf{K}_{t}$ and $\textbf{W}_{t}$. We propose a novel specification for $\textbf{K}_{t}$ and $\textbf{W}_{t}$ that provides an extension of the MI prior used by \citet{griffith2000,griffith2002,griffith2004}, \citet{griffith2007}, \citet{hughes}, and \citet{aaronp}. The MI prior for $\textbf{K}_{1}$ is given by $\textbf{K}_{1} = \sigma_{K}^{2} \textbf{S}_{X,1}^{\prime}\textbf{Q}_{1}\textbf{S}_{X,1}$, where $\sigma_{K}^{2}>0$ is unknown, $\textbf{Q}_{1} = \bm{1}_{N_{1}} - \textbf{A}_{1}$, and $\bm{1}_{N_{1}}$ is a $N_{1}$-dimensional vector of 1s. Notice that the MI prior is defined in the spatial only setting where $t = 1 = T$. Hence, we extend this prior to the multivariate-spatio-temporal setting. Moreover, we show that this extension allows one to incorporate knowledge of the spatial process.

Now, a reasonable criticism of the MI basis function is that we may be ignoring important sources of variability by restricting $\{\textbf{S}_{t}\}$ to a column space that is linearly independent of $\{\textbf{X}_{t}\}$; that is, for each $t$ requiring $\textbf{S}_{t}\in \mathcal{C}(\textbf{P}_{X,t})$, where $\textbf{P}_{X,t}\equiv \textbf{X}_{t}(\textbf{X}_{t}^{\prime}\textbf{X}_{t})^{-1}\textbf{X}_{t}$ and the column space of $\textbf{P}_{X,t}$ is denoted as $\mathcal{C}(\textbf{P}_{X,t})$. To see this, rewrite (\ref{process:model}) and let $\textbf{S}_{t} = [\textbf{H}_{X,t}, \textbf{L}_{X,t}]$ and $\bm{\eta}_{t}\equiv (\bm{\kappa}_{X,t}^{\prime},\bm{\delta}_{X,t})^{\prime}$ so that
\begin{align}\label{decomp2}
\by_{t} &= \textbf{X}_{t}\bm{\beta}_{t} + \textbf{H}_{X,t}\bm{\kappa}_{X,t} + \textbf{L}_{X,t}\bm{\delta}_{X,t}+ \bm{\xi}_{t}; \hspace{5pt}t = 2,\ldots,T.
\end{align}
\noindent
Here, the $N_{t}\times h$ matrix $\textbf{H}_{X,t}\in \mathcal{C}(\textbf{P}_{X,t})$, the $N_{t}\times l$ matrix $\textbf{L}_{X,t}\in \mathcal{C}(\textbf{P}_{X,t})^{\perp}$, $h$ and $l$ are non-negative integers, $\bm{\kappa}_{X,t}$ is a $h$-dimensional Gaussian random vector, and $\bm{\delta}_{X,t}$ is a $l$-dimensional Gaussian random vector; $t = 2,\ldots,T$. The decomposition in (\ref{decomp2}) is the space-time analogue of the decomposition used for discussion in \citet{Reich} and \citet{hughes}. The use of MI basis functions is equivalent to setting $h$ equal to $r$, $\textbf{H}_{X,t} = \textbf{S}_{X,t}$, and $\textbf{L}_{X,t}$ equal to a $n_{t}\times l$ matrix of zeros for each $t$. As a result, the model based on MI basis functions ignores the variability due to $\{\bm{\delta}_{X,t}\}$. In a similar manner, one can argue that the MI propagator matrix may also ignore other sources of variability.

To address this concern we consider specifying $\{\textbf{K}_{t}\}$ as positive semi-definite matrices that are ``close'' to target precision matrices (denoted with $\textbf{P}_{t}$ for $t = 1,\ldots,T$) that do not ignore these sources of variability; critically, the use of a target precision matrix allows us to reduce the parameter space. Then, let $\textbf{K}_{t} = \sigma_{K}^{2}\textbf{K}_{t}^{*}(\textbf{P}_{t})$, where $\sigma_{K}^{2}>0$ is unknown and
\begin{equation}\label{Kstar}
 \textbf{K}_{t}^{*}(\textbf{P}_{t})=\underset{\textbf{C}}{\mathrm{arg\hspace{5pt}min}}\left\lbrace ||\textbf{P}_{t} - \textbf{S}_{X,t}\textbf{C}^{-1}\textbf{S}_{X,t}^{\prime}||_{\mathrm{F}}^{2}\right\rbrace;\hspace{5pt} t = 1,\ldots,T.
 \end{equation}
 \noindent
 Here, $||\cdot||_{\mathrm{F}}$ denotes the Frobenius norm. In (\ref{Kstar}), we minimize the Frobenius norm across the space of positive semi-definite matrices. In a similar manner, if $\textbf{P}_{t}\equiv \textbf{P}$, $\textbf{X}_{t}\equiv \textbf{X}$, $\textbf{S}_{X,t}\equiv \textbf{S}_{X}$, $\textbf{M}_{B,t}\equiv \textbf{M}_{B}$,  $\textbf{K}_{t}\equiv \textbf{K}$, and $\textbf{W}_{t}\equiv \textbf{W}$ then we let $\textbf{K} = \sigma_{K}^{2}\textbf{K}^{*}(\textbf{P})$, where $\sigma_{K}^{2}>0$ is unknown and
\begin{equation}\label{Kstarnot}
 \textbf{K}^{*}(\textbf{P})=\underset{\textbf{C}}{\mathrm{arg\hspace{5pt}min}}\left\lbrace ||\textbf{P} - \textbf{S}_{X}\textbf{C}^{-1}\textbf{S}_{X}^{\prime}||_{\mathrm{F}}^{2}\right\rbrace.
 \end{equation}
\noindent 	 
In the following proposition we show how to compute $\textbf{K}_{t}^{*}$ in (\ref{Kstar}) for $t = 1,\ldots,T$ and $\textbf{K}^{*}$ in (\ref{Kstarnot}).\\
 
 \noindent
 \textit{Proposition 1:} Let $\bm{\Phi}_{k}$ be a generic $n \times r$ real matrix such that $\bm{\Phi}_{k}^{\prime}\bm{\Phi}_{k}=\textbf{I}_{r}$, $\textbf{C}$ be a generic $r \times r$ positive definite matrix, $\textbf{P}_{k}$ be a generic $n\times n$ positive definite matrix, and let $k = 1,\ldots,K$. Then, the value of $\textbf{C}$ that minimizes $\sum_{k=1}^{K}||\textbf{P}_{k} - \bm{\Phi}_{k}\textbf{C}^{-1}\bm{\Phi}_{k}^{\prime}||_{F}^{2}$ within the space of positive semi-definite covariances is given by,
 \begin{equation}\label{minfrobs}
 \textbf{C}^{*} = \left\lbrace\mathcal{A}^{+}\left(\frac{1}{K}\sum_{k = 1}^{K} \bm{\Phi}_{k}^{\prime}\textbf{P}_{k}\bm{\Phi}_{k}\right)\right\rbrace^{-1},
 \end{equation}
 \noindent
 where $\mathcal{A}^{+}$($\textbf{R}$) is the best positive approximate \citep{Higham} of a real square matrix $\textbf{R}$. Similarly, the value of $\textbf{C}$ that minimizes $\sum_{k=1}^{K}||\textbf{P}_{k} - \bm{\Phi}_{k}\textbf{C}\bm{\Phi}_{k}^{\prime}||_{F}^{2}$ within the space of positive semi-definite covariances is given by,
 \begin{equation}\label{minfrobs2}
 \mathcal{A}^{+}\left(\frac{1}{K}\sum_{k = 1}^{K} \bm{\Phi}_{k}^{\prime}\textbf{P}_{k}\bm{\Phi}_{k}\right).
 \end{equation}
 \noindent
 \textit{Proof:} See {Appendix A}. \\
 
 \noindent
 If we let $K=T$, $\bm{\Phi}_{k}=\textbf{S}_{X,k}$ for each $k$, then the corresponding expression of (\ref{minfrobs}) yields $\textbf{K}_{t}^{*}$ in (\ref{Kstar}). Likewise, if we let $K=1$, $\bm{\Phi}_{1}=\textbf{S}_{X}$, then the corresponding expression of (\ref{minfrobs}) yields $\textbf{K}^{*}$ in (\ref{Kstarnot}).
 
 With both $\{\textbf{K}_{t}\}$ and $\{\textbf{M}_{t}\}$ specified we can solve for $\{\textbf{W}_{t}\}$. That is, using the VAR(1) model
 \begin{equation}\label{Wstar}
 \textbf{W}_{t} = \textbf{K}_{t} - \textbf{M}_{B,t}\textbf{K}_{t-1}\textbf{M}_{B,t}^{\prime}\equiv \sigma_{K}^{2}\textbf{W}_{t}^{*}; \hspace{5pt} t = 2,\ldots,T,
 \end{equation}
\noindent
or
 \begin{equation}\label{Wstarnot}
 \textbf{W} = \textbf{K} - \textbf{M}_{B}\textbf{K}\textbf{M}_{B}^{\prime}\equiv \sigma_{K}^{2}\textbf{W}^{*},
 \end{equation}
 \noindent
 in the case where $\textbf{X}_{t}\equiv \textbf{X}$, $\textbf{S}_{X,t}\equiv \textbf{S}_{X}$, $\textbf{M}_{B,t}\equiv \textbf{M}_{B}$,  $\textbf{K}_{t}\equiv \textbf{K}$, and $\textbf{W}_{t}\equiv \textbf{W}$. In (\ref{Wstar}) and (\ref{Wstarnot}), the $r\times r$ matrices $\textbf{W}_{t}^{*} = \textbf{K}_{t}^{*} - \textbf{M}_{B,t}\textbf{K}_{t-1}^{*}\textbf{M}_{B,t}^{\prime}$ and $\textbf{W}^{*} = \textbf{K}^{*} - \textbf{M}_{B}\textbf{K}^{*}\textbf{M}_{B}^{\prime}$; $t = 2,\ldots,T$. It is important to note that the $r\times r$ matrices in the set $\{\textbf{W}_{t}^{*}\}$ (or the $r\times r$ matrix $\textbf{W}^{*}$) may not necessarily positive semi-definite. If $\textbf{W}_{t}^{*}$ is not positive semi-definite for some $t$ then we suggest using the best positive approximate. This is similar to ``lifting'' adjustments suggested by \citet{kang-cressie-shi-2010} in the spatio-temporal setting.
 
There are many choices for the ``target precision'' matrices $\{\textbf{P}_{t}\}$ in (\ref{Kstar}) and (\ref{Wstar}). For example, one might use the CAR model and let $\textbf{P}_{t} = \textbf{Q}_{t}$, where $\textbf{Q}_{t} = \bm{1}_{N_{t}} - \textbf{A}_{t}$ and $\bm{1}_{N_{t}}$ is a $N_{t}$-dimensional vector of 1s; $t = 1,\ldots,T$. This allows one to incorporate neighborhood information into the priors for $\textbf{K}$ and $\textbf{W}$. In the case where the areal units are small and regularly spaced, one might consider the many spatio-temporal covariance functions that are available (e.g., see  \citet{Gneitingcorr}, \citet{HuangCressie2}, and \citet{steinSep}). An empirical Bayesian approach might be considered and an estimated precision (or covariance) matrix might be used (e.g., see \citet{guttorpandpeterson}).

An additional motivation for the restrictions in (\ref{Kstar}) and (\ref{Kstarnot}) is that the MI prior can be interpreted as a special case. This is formally stated in Corollary 1 below.\\

 \noindent
 \textit{Corollary 1:} Let $\textbf{S}_{X,1}$ be the MI propagator matrix, and $\textbf{C}$ be a generic $r \times r$ positive definite matrix. Then, the value of $\textbf{C}$ that minimizes $||\textbf{Q}_{1} - \textbf{S}_{X,1}\textbf{C}\textbf{S}_{X,1}^{\prime}||_{F}^{2}$ within the space of positive semi-definite covariances is given by,
 \begin{equation}\label{minfrobscor}
 \textbf{C}^{*} = \mathcal{A}^{+}\left(\textbf{S}_{X,1}^{\prime}\textbf{Q}_{1}\textbf{S}_{X,1}\right).
 \end{equation}
 \noindent
 \textit{Proof:} Let $K = 1$, $\bm{\Phi}_{1} = \textbf{S}_{X,1}$, and $\textbf{P}_{1} = \textbf{Q}_{1}$. Then, apply Proposition 1.\\
 
 \noindent
 If $\textbf{S}_{X,1}^{\prime}\textbf{Q}_{1}\textbf{S}_{X,1}$ is positive definite, then (\ref{minfrobscor}) is equal to the MI prior. \citet{aaronpBayes} show that $\textbf{S}_{X,1}^{\prime}\textbf{Q}_{1}\textbf{S}_{X,1}$ is positive definite as long as an intercept is included in the definition of $\textbf{X}_{1}$.

The prior distributions for these parameters are specified to be Gaussian (for $\{\bm{\beta}_{t}\}$) and inverse gamma (IG) priors (for $\sigma_{K}^{2}$, $\sigma_{W}^{2}$, and $\{\sigma_{\xi,t}^{2}\}$). This will allow us to use conjugacy to obtain exact expressions for the full-conditionals within a Gibbs sampler. See {Appendix B} for the details regarding the MCMC algorithm.

\section{Applications}\label{sec:app} To illustrate the variety of random processes that can be modeled using the MSTM, we consider three important problems. In doing so, we demonstrate that the MSTM provides an extremely viable solution to many important problems found in different corners of the spatial statistics literature.

The first problem is modeling temporal nonstationarity. This is a long-standing problem within the time-series literature \citep[e.g., see][]{Dahlhaus}, and we extend it to the multivariate-spatio-temporal setting. This is demonstrated in Section~\ref{sec:cancer}, where we consider US cancer mortality data. 

The second problem is multivariate-spatio-temporal prediction (by prediction we mean estimating latent random processes) using data from multiple surveys. The MSTM is flexible enough to solve this problem; here, we only need to set $\{v_{t}^{(\ell)}(\cdot)\}$ (i.e., the variance of $\{\epsilon_{t}^{(\ell)}\}$) equal to the {survey variance provided by each of the available statistical agencies}. In general, there are many models used to combine surveys in the time-series setting \citep[e.g., see][]{Jones,Bell,Feder}, but to our knowledge nothing has been proposed that would be suitable for the multivariate-spatio-temporal data setting. In fact, much of the literature involves simplifications that are not appropriate for our setting; for example, \citet{Merkouris1} and \citet{dorfman} assume marginal independence between surveys and \citet{Elliott}, \citet{Keller}, and \citet{lohr} consider a dual frame design in the non-spatio-temporal setting. \citet{wang} and \citet{Giorgi} consider a fully Bayesian approach to combine surveys in the time series and spatial settings, respectively. For other examples of Bayesian approaches to combining multiple surveys see \citet{Raghunathan}, \citet{Bryant}, and the references therein.  Although there has been considerable methodological development for combining data from multiple surveys, none of the previously mentioned approaches incorporate multivariate spatio-temporal dependencies. As such, in Section~\ref{sec:comb}, we use the MSTM in a fully Bayesian framework to combine unemployment rates from ACS and LAUS.

In practice, multivariate-spatio-temporal datasets can be extremely large. Hence, the third problem we consider is spatial prediction using massive datasets. Spatial datasets are becoming increasingly larger and, as a result, new methodologies {(that are not directly applicable in the multivariate-spatio-temporal setting)} are being proposed to address the computational bottleneck involved with spatial prediction using large spatial datasets. Specifically, the Gaussian likelihood involves the computation of an inverse and a determinant of a $n\times n$ covariance matrix; a task that is on the order of $n^{3}$ computations. The reduced rank structure that we impose in (\ref{process:model}) allows us to avoid computing the likelihood of a large dimensional Gaussian random vector if $r \ll n$. In Section~\ref{sec:massive}, we analyze a massive multivariate-spatio-temporal dataset consisting of average monthly income estimates obtained from the US Census Bureau's {LEHD} survey.

For all three of the following examples, the Gibbs sampler, provided in Appendix~B, was run for 10,000 iterations with a burn-in of 1,000 iterations. Convergence of the {Markov chain Monte Carlo algorithm}  was assessed visually using trace plots of the sample chains, with no lack of convergence detected.

\subsection{An Analysis of U.S. Cancer Mortality Rates}\label{sec:cancer} The NCI is a division of the National Institutes of Health (NIH). They generate a {wide range} of cancer statistics recorded over the US, which we use to provide an analysis of mortality rates due to cancer (per 100$\--$thousand) by gender (see, http://statecancerprofiles.cancer.gov/). These mortality rates are recorded over each state from 1975 to 2010; that is, let $\ell = 1$ indicate females, $\ell = 2$ indicate males, $D_{\mathrm{O},t}^{(\ell)}$ consists of each state in the US, $T_{L}^{(\ell)} = 1$, and $T_{U}^{(\ell)} = 35$ (for notational convenience we shift the time period from 1975$\--$2010 to 1$\--$35). In Figures 1(a) and 1(b), we provide selected maps of the mortality rates as estimated by NCI. Here, we see that in general NCI estimates higher cancer mortality rates in the east coast than in the west coast for both women and men.

The primary goals of our analysis are: to specify the MSTM in a way that allows for the complex dependencies found in US cancer mortality rates; to estimate cancer mortality rates; and to interpret estimates from the MSTM. Recent studies have compared cancer mortality forecasts based on methods with both stationary
and nonstationary models in time \citep{cancertime,cancertime2}. These studies suggest that the assumption of nonstationarity in time is reasonable and that one would also expect interactions between space and time. Consequently, we specify the MSTM in (\ref{data:model}) and (\ref{process:model}) to allow for these properties.

To obtain nonstationarity in time we require the MI propagator matrix to differ at different time points. Upon inspection of the definitions of $\{\textbf{S}_{X,t}\}$ and $\{\textbf{M}_{B,t}\}$ we see that this can be achieved by letting $\textbf{x}_{t}^{(\ell)}$ change over time. Hence, for illustrative purposes we make the following specifications. Let $\textbf{x}_{t}^{(\ell)}(A)\equiv (1,I(\ell = 2),c_{1}, c_{2}, t, I(\ell = 2)\times c_{1}, I(\ell = 2)\times c_{2})^{\prime}$ be the 7$\--$dimensional vector of known covariates, where $\textbf{c} \equiv (c_{1}, c_{2})^{\prime}$ are the x$\--$y coordinates of the centroid of the generic areal unit $A$. Notice that we allow for two-way interactions between gender so that the changes in $\mu_{t}^{(\ell)}(\cdot)$ per unit increase in time and $\textbf{c}$ are different for each gender. Additionally, we let $r=12$, which is roughly 10$\%$ of the available MI basis functions at each time point $t$. Let $\{\textbf{Q}_{t}\}$, defined below (\ref{Wstar}), be the target precision matrix.

Preliminary analyses ({QQ plots} and histograms) indicate that the assumption of normality appears reasonable. Hence, we apply the MSTM to the NCI cancer mortality rates and in Figures 1(c)$\--$1(f), we provide selected maps of the predicted cancer mortality rates and their corresponding posterior mean squared prediction error (MSPE) for women and men, respectively. Here, we see that in general there are higher cancer mortality rates in the east coast than in the west coast. For males, this pattern is more apparent. 

Next, in Figure~2 we plot the posterior mean of the regression parameter corresponding to time (i.e., $\beta_{3,t}$, where $\bm{\beta}_{t}=(\beta_{1,t},...,\beta_{7,t})^{\prime}$). Recall that the presence of time-varying covariates implies nonstationarity in time for the MSTM. Hence, if we observe a pattern that suggests that $\beta_{3,t}\equiv 0$ then the assumption of nonstationarity in time may not be reasonable. We see in Figure~2 that the mean cancer mortality rate decreases exponentially over time. Additionally, the 95$\%$ {pointwise credible intervals} do not contain zero at any time point, which suggests that nonstationarity in time is a reasonable assumption.

\subsection{Combining Missouri Unemployment Rates From ACS and LAUS}\label{sec:comb} {The Bureau of Labor Statistics administers the LAUS program}, which provides unemployment statistics recorded over the US. LAUS defines unemployment as all out-of-work individuals who are available to have a job and have sought work in the past four weeks of taking the survey. In this section, we provide an analysis of these unemployment rates (in percent) recorded over each county in the state of Missouri from 1990 to 2012; that is, $D_{\mathrm{O},t}^{(1)}$ consists of Missouri counties, $T_{L}^{(1)} = 1$, and $T_{U}^{(1)} = 23$ (for notational purposes we shift the time period from 1990$\--$2012 to 1$\--$23).

However, LAUS is not the only data source available for unemployment rates over the US. In particular, ACS also provides 1$\--$year period estimates of the unemployment rates from 2005 to 2012 over counties in Missouri (see http://factfinder2.census.gov/). ACS shares the same definition of unemployment as LAUS. In addition to unemployment rates, we also consider a demographic variable that is related to unemployment rates; specifically, we analyze 1$\--$year ACS period estimates of median household income ($\ell=2$) from 2005 to 2012. In Figures 3(a)$\--$3(c), we provide selected maps of the LAUS unemployment rate, and the unemployment rate and median income as estimated by ACS. The ACS estimates have considerably less spatial coverage than the LAUS estimates, since ACS does not provide 1$\--$year period estimates for every county of Missouri.  Additionally, ACS estimates are available across multiple variables, which is not the case for LAUS.

The primary goals of our analysis are to estimate and interpret the unemployment rates and determine whether or not combining the estimates from LAUS and ACS leads to a substantial improvement in the estimates. One difficulty with interpreting ACS estimates is that they often have large margins of error \citep{speilman}. The fact that ACS estimates can have large sampling variability makes it an interesting example in that the incorporation of LAUS may lead to more precise estimates of Missouri unemployment rates.

Preliminary analyses using {QQ plots} and histograms indicate that the logit (log) of the unemployment rates (median income) is roughly Gaussian. Since we assume that the underlying data is Gaussian we treat the logit (log) of the unemployment rates (median income) as $\{Z_{t}^{(\ell)}(\cdot)\}$ in (\ref{data:model}). That is, let $R_{t}^{(\ell)}(\cdot)$ represent the estimates from ACS and LAUS; $\ell = 1,2$ and $t = 1,...,23$. The logit transformation is given by $\mathrm{logit}(w) \equiv \mathrm{log}\{w/(1-w)\}$ for a generic real number $w\in [0,1]$. Then, we define $Z_{t}^{(1)}(\cdot) \equiv \mathrm{logit}(R_{t}^{(1)}(\cdot))$ and $Z_{t}^{(2)}(\cdot) \equiv \mathrm{log}(R_{t}^{(1)}(\cdot))$ for $t = 1,...,23$. The survey variance estimates are approximated on the transformed scale using the Delta-method (e.g., see \citet{delta}). 

For illustrative purposes we make the following specifications. Set the target precision matrices equal to $\{\textbf{Q}_{t}\}$ as previously described below (\ref{Wstar}). Let $\textbf{x}_{t}^{(1)}(A)\equiv (1,c_{1}, c_{2})^{\prime}$, where we recall $(c_{1}, c_{2})^{\prime}$ are the x$\--$y coordinates of the centroid of the generic areal unit $A$. Also, let $r=10$, which is roughly 10$\%$ of the available MI basis functions at each time point $t$. In Figures 3(d) and 3(e) we present a selected map of the predicted unemployment rate and the associated estimates of MSPE of the unemployment rate (on the original scale). The values of MSPE at each location is small (on the order of $10^{-7}$); thus, we appear to be obtaining precise estimates of the hidden process. In Figure 4, we plot $(1/|D_{\mathrm{P},t}^{(1)}|)\times \sum_{A}\widehat{Y}_{t}^{(1)}(A)$ versus year $t$, where $\widehat{Y}_{t}^{(\ell)}(\cdot)$ denotes the predictions based on the MSTM. This plot shows a decrease in the unemployment rate until the early 2000s followed by an increase. Then, in 2007 there was a sharp increase in the unemployment rate until 2010. This conforms to intuition since 2007 marks the start of the US housing crisis \citep[e.g., see][]{holt}.

{Now, denote the multivariate spatial predictors based solely on ACS (LAUS) estimates with $\widehat{Y}_{t}^{(\ell,1)}$ ($\widehat{Y}_{t}^{(\ell,2)}$); $t = 1,...,23$ and $\ell = 1,...,T$. Consider the relative leave-one survey-out (RLS) criterion
\begin{equation*}
\mathrm{RLS}(m)\equiv \frac{\sum_{j = 1}^{J}\sum_{t = 1}^{T}\underset{A \in D_{\mathrm{O},t}^{(\ell)}}{\sum}(Y_{t}^{(1)}(A; \bz)^{[j]} - \widehat{Y}_{t}^{(1,m)}(A))^{2}}{\sum_{j = 1}^{J}\sum_{t = 1}^{T}\underset{A \in D_{\mathrm{O},t}^{(1)}}{\sum}(Y_{t}^{(1)}(A; \bz)^{[j]} - \widehat{Y}_{t}^{(1)}(A))^{2}};\hspace{5pt} m = 1,2,
\end{equation*}
where $Y_{t}^{(\ell)}(A; \bz)^{[j]}$ represents the $j$$\--$th MCMC replicate of $Y_{t}^{(\ell)}(A)$ computed using the entire $n$$-$dimensional data-vector $\bz \equiv (Z_{t}^{(\ell)}(A): \ell = 1,...,L, t = T_{L}^{(\ell)},...,T_{U}^{(\ell)}, A \in D_{\mathrm{P},t}^{(\ell)})^{\prime}$ and $j = 1,...,J$. That is, $\mathrm{RLS}(m)$ is the MSPE of $\widehat{Y}_{t}^{(1)}$ relative to the MSPE of $\widehat{Y}_{t}^{(1,m)}$ for survey $m$. Values of RLS that are smaller than 1 indicate that combining surveys does not lead to an improvement in MSPE, while values larger than 1 indicate an improvement in MSPE. Also, if $1<\mathrm{RLS}(2)< \mathrm{RLS}(1)$ then this indicates that Survey 1 benefits more from combining surveys (in terms of reducing MSPE) than Survey 2. For this example, $\mathrm{RLS}(2) = 3.03\times 10^5$ and $\mathrm{RLS}(1) = 3.01\times 10^7$, which is considerably larger than 1. Hence, we see a dramatic improvement in the MSPE when using both surveys as opposed to using a single survey. Additionally, we see that ACS benefits more from combining surveys than LAUS, since $1<\mathrm{RLS}(2)< \mathrm{RLS}(1)$.}

\subsection{Predicting Average Monthly Income Using a Massive Dataset}\label{sec:massive} We demonstrate the use of MSTM using a massive multivariate-spatio-temporal dataset
made up of data obtained from the LEHD program, which is administered by the US Census Bureau. The LEHD program provides public-use survey data by combining Census Bureau survey data on employers and employees. Public access data on several earnings and other economic variables are available quarterly on various geographies of the US (see, http://www.census.gov/ces/dataproducts/). 

In this section, we consider the average monthly income by individuals with steady jobs for each quarter from 1990 to 2013 ($T = 92$), and by US counties ($\{D_{\mathrm{P},t}^{(\ell)}\}$ and $\{D_{\mathrm{O},t}^{(\ell)}\}$ both consist of US counties). These income estimates are available by industry (see Appendix~C for a list) and gender. Each industry/gender combination identifies a unique multivariate-spatio-temporal field; hence, $\ell = 1,...,L = 40$. In total, there are 7,530,037 observations over the entire US in this dataset, which we jointly analyze using the MSTM. We present a subset of this dataset in Figures 5(a) and 5(b). We see that the average monthly income is relatively constant across each county of the state of Missouri, and that men tend to have higher average monthly income than women. This pattern is consistent across the different spatial locations, industries, and time-points.

The primary goals of our analysis are to estimate the average monthly income, and determine whether or not it is computationally feasible to use the MSTM for a dataset of this size. Preliminary analyses indicate that the log average monthly income is roughly Gaussian. Since we assume that the underlying data is Gaussian we treat the log of the average income as $\{Z_{t}^{(\ell)}(\cdot)\}$ in (\ref{data:model}). The survey variance estimates are approximated on the transformed scale using the Delta-method (e.g., see \citet{delta}). 

For illustration, we make the following specifications. Set the target precision matrix equal to $\{\textbf{Q}_{t}\}$ as previously described below (\ref{Wstar}). Let $\textbf{x}_{t}^{(\ell)}(A)\equiv (1,I(\ell = 1),...,I(\ell = 39), I(g = 1)\times I(\ell = 1),...,I(g = 1)\times I(\ell = 39))^{\prime}$, where $g = 1,2$ indexes males and females, respectively, and recall $I(\cdot)$ is the indicator function. Also, let $r=20$, which is roughly 30$\%$ of the available MI basis functions at each time point $t$. Using the MSTM with these specifications we predict $L\times T = 40\times 92 = 3,680$ different spatial fields. The CPU time required to compute these predictions is approximately 2.3 days, and all of our computations were performed on a dual 10 core 2.8 GHz Intel Xeon E5-2680 v2 processor, with 256 GB of RAM. Of course, additional efforts in efficient programming may result in faster computing; however, these results indicate that it is practical to use the MSTM to analyze massive data from a computational point of view.

Although we modeled the entire US simultaneously, for illustration, we present maps of predicted monthly income for the state of Missouri, for each gender, for the education industry, and for the 92-nd quarter (Figures~5(c) and 5(d)). The prediction maps are essentially constant over the state of Missouri, where women tend to have a predicted monthly income of slightly less than 1,200 dollars and men consistently have a predicted monthly income of about 1,800 dollars. As observed in Figure~5(a) and (b), there is a clear pattern where men have higher predicted monthly income than women. These predictions appear reasonable since the maps of the posterior square root MSPE, in Figures~5(e) and 5(f), indicate we are obtaining precise predictions. Additionally, upon comparison of Figures~5(a) and 5(b) to Figures 5(c) and 5(d), we see that the predictions reflect the same general pattern in the data. These results are similar across the different states, industries, and time-points.

To further corroborate the patterns in the MSTM predictions we fit the univariate spatial model from \citet{hughes}, which is currently the alternative model for spatial prediction of large areal datasets. We fit the univariate spatial model from \citet{hughes} to the data in Figures 5(a) and 5(b) with $r=62$ basis functions ($100\%$ of the available basis functions) and obtain the prediction maps (not shown). Notably, the predictions are also fairly constant around 1,200 and 1,800 dollars. Moreover, the MSPE of the \citet{hughes} predictions (summed over all US counties) is 3.23 times larger than the MSPE of the predictions from the MSTM summed over all US counties. This may be due, in part, to the fact that the model in \citet{hughes} does not incorporate dependencies that arise from different variables and time-points.

It should be mentioned that, despite the inherent computational issues, having an abundance of data has advantages. For example, notice in Figure~5(b) that LEHD does not release data at two counties of Missouri for men in the education industry during quarter 92. Although these values are missing for this variable and time-point, LEHD releases data at these two counties (for men in the education industry) for 43 different quarters. Hence, with the observed values from 43 different spatial fields, we reduce the variability of predictions at the two missing counties during the 92-nd quarter (compare Figure~5(b) to 5(f)).

\section{Discussion}\label{sec:disc}  We have introduced fully Bayesian methodology to analyze areal datasets with multivariate-spatio-temporal dependencies. In particular, we introduce the multivariate-spatio-temporal mixed effects model (MSTM). {To date, little} has been proposed to model areal data that exhibit multivariate-spatio-temporal dependencies; furthermore, the available alternatives (see, \citet{carlinmst} and \citet{daniels}) do not allow for certain complexities in cross-covariances. Hence, the MSTM provides an important addition to the multivariate-spatio-temporal literature.

Our modeling decisions are made in an effort to allow the MSTM to be applied to a wide array of datasets. For example, we use a reduced rank approach to allow for massive multivariate-spatio-temporal datasets. Additionally, we allow for nonstationary and nonseparable multivariate-spatio-temporal dependencies, which is appropriate for many settings. This is achieved, in part, through a novel propagator matrix for a first-order vector autoregressive (VAR(1)) model, which we call the MI propagator matrix. This propagator matrix is an extension of the MI basis function \citep{griffith2000, griffith2002, griffith2004, griffith2007,hughes,aaronp} from the spatial only setting to the multivariate-spatio-temporal setting. We motivate both the MI basis function and the MI propagator matrix as an approximation to a target precision matrix, that allows for (1) computationally efficient statistical inference and (2) identifiability of regression parameters.

We also make an effort to allow practitioners to incorporate knowledge of the spatial process into the MSTM. Specifically, we propose an extension of the MI prior to the spatio-temporal case. This extension shows that the covariance matrix of the random effect is close (in Frobenius norm) to a ``target precision'' matrix, which is chosen based on knowledge of the underlying spatial process. In general, this contribution has implications for defining informative parameter models for high-dimensional spatio-temporal processes.

To demonstrate the effectiveness and broad applicability of our approach, we consider three motivating examples. In the first application, we analyze US cancer mortality rates using the MSTM. Here, nonstationary and nonseparablity are realistic assumptions \citep{cancertime,cancertime2}, which can easily be incorporated into the MSTM. Estimates of the MSPE indicate that the proposed BHM leads to precise predictions and estimates. Furthermore, the results of this study suggest that the assumption of nonstationarity in time is reasonable.

In the second example, we consider combining {data from multiple repeated surveys}, which is a topic of general interest. To demonstrate this, we consider unemployment rates from both ACS and LAUS. Here, the MSTM is used to combine ACS and LAUS estimated unemployment rates in Missouri. Estimates of the MSPE indicate that combining these surveys using the proposed BHM leads to a more precise estimate of unemployment rate than using each survey individually.

In the third example we consider a massive dataset of monthly income. The dataset consists of 7,530,037 observations, which is used to predict 3,680 different spatial fields consisting of all the counties in the US. The recorded CPU time for this example was 2.3 days, which indicates that it is reasonable to use the MSTM for massive data.

There are many opportunities for future research. In particular, the parameter model introduced in Section~\ref{sec:parmod} is of independent interest. In our applications, we let $\{\textbf{Q}_{t}\}$ give the target precision. However, one could conceive of many different ``target precisions'' built from deterministic models, for say, atmospheric variables. Another avenue {for} future research is to develop the MI propagator matrix, which was only provided for VAR(1). One could easily use this strategy for other time series models.

\section*{Acknowledgments} This research was partially supported by the U.S. National Science Foundation (NSF) and the U.S. Census Bureau under NSF grant SES$\--$1132031, funded through the NSF-Census Research Network (NCRN) program.

\section*{Appendix A: The Proof of Proposition 1}
\renewcommand{\theequation}{A.\arabic{equation}}
\setcounter{equation}{0}
By definition of the Frobenius norm
\begin{align}\label{frobnormproof}
\nonumber
&\sum_{k = 1}^{K}||\textbf{P}_{k} - \bm{\Phi}_{k}\textbf{C}^{-1}\bm{\Phi}_{k}^{\prime}||_{F}^{2} = \sum_{k = 1}^{K}\mathrm{trace}\left\lbrace\left(\textbf{P}_{k} - \bm{\Phi}_{k}\textbf{C}^{-1}\bm{\Phi}_{k}^{\prime}\right)^{\prime}\left(\textbf{P}_{k} - \bm{\Phi}_{k}\textbf{C}^{-1}\bm{\Phi}_{k}^{\prime}\right)\right\rbrace \\
\nonumber
&= \sum_{k = 1}^{K}\left\lbrace\mathrm{trace}\left(\textbf{P}_{k}^{\prime}\textbf{P}_{k}\right) - 2\times \mathrm{trace}\left( \bm{\Phi}_{k}^{\prime}\textbf{P}_{k}^{\prime}\bm{\Phi}_{k}\textbf{C}^{-1}\right) + \mathrm{trace}\left(\textbf{C}^{-2}\right)\right\rbrace\\
\nonumber
& = \sum_{k = 1}^{K}\mathrm{trace}\left(\textbf{P}_{k}^{\prime}\textbf{P}_{k}\right)- K\times \mathrm{trace}\left\lbrace \left(\frac{1}{K}\sum_{k = 1}^{K}\bm{\Phi}_{k}^{\prime}\textbf{P}_{k}\bm{\Phi}_{k}\right)^{2}\right\rbrace\\
&+K\times||\textbf{C}^{-1} - \frac{1}{K}\sum_{k = 1}^{K}\bm{\Phi}_{k}^{\prime}\textbf{P}_{k}\bm{\Phi}_{k}||_{F}^{2}.
\end{align}
\noindent
It follows from Theorem 2.1 of \citet{Higham} that the minimum of (\ref{frobnormproof}) is given by Equation (10) in the main document. In a similar manner, if one substitutes $\textbf{C}$ for $\textbf{C}^{-1}$ in (\ref{frobnormproof}) then we obtain the result in Equation (11) in the main document.

\section*{Appendix B: Full-Conditionals for the Gibbs Sampler} 
\renewcommand{\theequation}{B.\arabic{equation}}
\setcounter{equation}{0}
The model that we use for multivariate-spatio-temporal data is given by:
\begin{align}\label{summary}
\nonumber
&\mathrm{Data\hspace{5pt}Model:}\hspace{5pt}Z_{t}^{(\ell)}(A)\vert \bfbeta_{t}, \bm{\eta}_{t}, \xi_{t}^{(\ell)}(\cdot)\ind \mathrm{Normal}\left(\textbf{x}_{t}^{(\ell)}(A)^{\prime}\bfbeta_{t} + \textbf{S}_{X,t}^{(\ell)}(A)^{\prime}\bm{\eta}_{t} + \xi_{t}^{(\ell)}(A), v_{t}^{(\ell, m)}(A)\right);\\
\nonumber
&\mathrm{Process\hspace{5pt}Model\hspace{5pt}1:}\hspace{5pt} \bm{\eta}_{t}\vert \bm{\eta}_{t-1},\textbf{M}_{B,t},\textbf{W}_{t}\sim \mathrm{Gaussian}\left(\textbf{M}_{B,t}\bm{\eta}_{t-1}, \textbf{W}_{t}\right);\\
\nonumber
&\mathrm{Process\hspace{5pt}Model\hspace{5pt}2:}\hspace{5pt} \bm{\eta}_{1}\vert \textbf{K}_{1} \sim \mathrm{Gaussian}\left(\bm{0}, \textbf{K}_{1}\right);\\
\nonumber
&\mathrm{Process\hspace{5pt}Model\hspace{5pt}3:}\hspace{5pt} \xi_{t}^{(\ell)}(\cdot)\vert \sigma_{\xi,t}^{2} \ind \mathrm{independent\hspace{4pt}Normal}\left(0, \sigma_{\xi,t}^{2}\right);\\
\nonumber
&\mathrm{Parameter\hspace{5pt}Model\hspace{5pt}1:}\hspace{5pt} \bfbeta_{t} \sim \mathrm{Gaussian}\left(\bm{\mu}_{\beta}, \sigma_{\beta}^{2}\textbf{I}_{p}\right);\\
\nonumber
&\mathrm{Parameter\hspace{5pt}Model\hspace{5pt}2:}\hspace{5pt} \sigma_{\xi,t}^{2} \sim \mathrm{IG}\left(\alpha_{\xi}, \beta_{\xi}\right);\\
\nonumber
&\mathrm{Parameter\hspace{5pt}Model\hspace{5pt}3:}\hspace{5pt}  \sigma_{K,t}^{2} \sim \mathrm{IG}\left(\alpha_{K}, \beta_{K}\right);\hspace{5pt} \ell = 1,...,L, t = T_{L}^{(\ell)},...,T_{U}^{(\ell)},A \in D_{\mathrm{P},t}^{(\ell)},
\end{align}
\noindent
where $\sigma_{\beta}^{2}>0$, $\alpha_{\xi}>0$, $\alpha_{K}>0$, $\beta_{\xi}>0$, and $\beta_{K}>0$. In Section 4 the prior mean of $\bm{\mu}_{\beta}$ is set equal to a $p$$\--$dimensional zero vector, and the corresponding variance $\sigma_{\beta}^{2}$ is set equal to $10^{15}$ so that the prior on $\{\bm{\beta}_{t}\}$ is vague. In Section 4, we also specify $\alpha_{\xi}$, $\alpha_{K}$, $\beta_{\xi}$, and $\beta_{K}$ so that the prior distributions of $\sigma_{\xi,t}^{2}$ and $\sigma_{K,t}^{2}$ are vague. Specifically, we let $\alpha_{\xi}=\alpha_{K}=2$, and $\beta_{\xi}=\beta_{K}=1$; here, the IG(2,1) prior is interpreted as vague since it has infinite variance. 

We now specify the full-conditionals for the process variables (i.e., $\{\bm{\eta}_{t}\}$ and $\{\xi_{t}^{(\ell)}(\cdot)\}$) and the parameters (i.e., $\{\bm{\beta}_{t}\}$, $\{\sigma_{\xi,t}^{2}\}$, and $\sigma_{K}^{2}$).\\

\noindent
\textit{Full-Conditionals for Process Variables:} Let the $n_{t}$$\--$dimensional random vectors $ \bm{\bz}_{t} \equiv \left(Z_{t}^{(\ell)}(A): \ell\right.$ $\left. = 1,...,L, A \in D_{\mathrm{O},t}^{(\ell)}\right)^{\prime}$, $ \bm{\xi}_{t} \equiv \left(\xi_{t}^{(\ell)}(A): \ell = 1,...,L, A \in D_{\mathrm{O},t}^{(\ell)}\right)^{\prime}$, and the $n_{t}\times p$ matrix $\textbf{X}_{t} \equiv \left(\textbf{x}_{t}^{(\ell)}(A):\right.$ $\left.\ell = 1,...,L, m = 1,...,M^{(\ell)}, A \in D_{\mathrm{O},t}^{(\ell)}\right)^{\prime}$; $t = 1,...,T$. Then, we update the full-conditional for $\bm{\eta}_{1:T} \equiv \left(\bm{\eta}_{t}^{\prime}: t = 1,...,T\right)^{\prime}$ at each iteration of the Gibbs sampler using the Kalman smoother. We accomplish this by performing the following steps:
\begin{enumerate}
	\item Find the Kalman filter using the shifted measurements $\{\widetilde{\bz}_{t}: \widetilde{\bz}_{t} = \bz_{t}-\textbf{X}_{t}\bm{\beta}_{t}-\bm{\xi}_{t}\}$\citep{shumway,cart1994,schnatter94,cressie-wikle-book}. That is, for $t = 1,...,T$ compute
	\begin{enumerate}
		\item $\bm{\eta}_{t\vert t}^{[j]} \equiv E\left(\bm{\eta}_{t}\vert \widetilde{\bz}_{1:t}, \bm{\theta}_{t}^{[j]}\right)$
		\item $\bm{\eta}_{t\vert (t-1)}^{[j]} \equiv E\left(\bm{\eta}_{t}\vert \widetilde{\bz}_{1:(t-1)}, \bm{\theta}_{t}^{[j]}\right)$
		\item $\textbf{P}_{t\vert t}^{[j]} \equiv \mathrm{cov}\left(\bm{\eta}_{t}\vert \widetilde{\bz}_{1:t}, \bm{\theta}_{t}^{[j]}\right)$	
		\item $\textbf{P}_{t\vert (t-1)}^{[j]} \equiv \mathrm{cov}\left(\bm{\eta}_{t}\vert \widetilde{\bz}_{1:(t-1)}, \bm{\theta}_{t}^{[j]}\right)$,		
	\end{enumerate}
	\noindent
	where $\textbf{P}_{1\vert 1}^{[j]} = (\sigma_{K}^{[j]})^{2}\textbf{K}^{*}$ and $\bm{\theta}_{t}^{[j]}$ represents the $j$$\--$th MCMC draw of $\bm{\theta}_{t}$ and $\sigma_{K}^{2}$, respectively.
	\item Sample $\bm{\eta}_{T}^{[j+1]} \sim \mathrm{Gaussian}\left(\bm{\eta}_{T\vert T}^{[j]}, \textbf{P}_{T\vert T}^{[j]}\right)$.
	\item For $t = T-1,T-2,...,1$ sample\\$\bm{\eta}_{t}^{[j+1]} \sim \mathrm{Gaussian}\left(\bm{\eta}_{t\vert t}^{[j]} + \textbf{J}_{t}^{[j]}(\bm{\eta}_{t+1}^{[j]} - \bm{\eta}_{t+1\vert t}^{[j]}), \textbf{P}_{t\vert t}^{[j]} - \textbf{J}_{t}^{[j]}\textbf{P}_{t+1\vert t}^{[j]}(\textbf{J}_{t}^{[j]})^{\prime} \right)$,\\where $\textbf{J}_{t}^{[j]} \equiv \textbf{P}_{t\vert t}^{[j]} \textbf{M}_{t}^{\prime}(\textbf{P}_{t+1\vert t}^{[j]})^{-1}$.
\end{enumerate}
Notice that within each MCMC iteraction we need to compute Kalman filter and Kalman smoothing equations. This adds more motivation for reduced rank modeling; that is, if $r$ is large (i.e., if $r$ is close in value to $n$) this step is not computationally feasible.

The remaining process variable $\{\xi_{t}^{(\ell)}(\cdot)\}$ can also be computed efficiently \citep{ravishank}. The full conditional for $\{\xi_{t}^{(\ell)}(\cdot)\}$ is given by: $\bm{\xi}_{t} \sim \mathrm{Gaussian}\left(\bm{\mu}_{\xi,t}^{*},\bm{\Sigma}_{\xi.t}^{*}\right),$ where $\bm{\Sigma}_{\xi,t}^{*} \equiv \left(\textbf{V}_{t} + \sigma_{\xi}^{2}\textbf{I}_{N_{t}}\right)^{-1}$, $\bm{\mu}_{\xi,t}^{*} \equiv \bm{\Sigma}_{\xi}^{*}\times \textbf{V}_{t}^{-1}\times(\bz_{t}-\textbf{X}_{t}\bm{\beta}_{t}-\textbf{S}_{t}\bm{\eta}_{t})$, $\textbf{V}_{t} \equiv \mathrm{diag}\left(v_{t}^{(\ell)}(A): \ell = 1,...,L, A \in D_{\mathrm{O},t}^{(\ell)}\right)$, and $\textbf{S}_{t} \equiv
\left(\textbf{S}_{t}^{(\ell)}(A): \ell = 1,...,\right.$ $\left. L,A \in D_{\mathrm{O},t}^{(\ell)}\right)^{\prime}$; $t = 1,...,T$.\\

\noindent
\textit{Full-Conditionals for the Parameters:} Similar to the full-conditional for $\{\xi_{t}^{(\ell)}(\cdot)\}$ \citep{ravishank} we also have the following full-conditional for ${\bm{\beta}_{t}}$:
$\bm{\beta}_{t} \sim \mathrm{Gaussian}\left(\bm{\mu}_{\beta,t}^{*},\bm{\Sigma}_{\beta,t}^{*}\right)$, where $\bm{\Sigma}_{\beta,t}^{*} \equiv \left( \textbf{X}_{t}^{\prime}\textbf{V}_{t}^{-1}\textbf{X}_{t}+ \sigma_{\beta}^{-2}\textbf{I}_{p} \right)^{-1}$, and
$\bm{\mu}_{\beta,t}^{*} \equiv \bm{\Sigma}_{\beta}^{*}\times \textbf{X}_{t}^{\prime}\textbf{V}_{t}^{-1}(\bz_{t}-\bm{\xi}_{t}-\textbf{S}_{t}\bm{\eta}_{t})$; $t = 1,...,T$.

Finally, the exact form of the full-conditionals for $\sigma_{K}^{2}$, $\sigma_{W}^{2}$, and $\{\sigma_{\xi,t}^{2}\}$ can also be found in a straightforward manner. It follows that the full conditionals for $\sigma_{K}^{2}$, $\sigma_{W}^{2}$, and $\sigma_{\xi,t}^{2}$ are IG($Tr/2 + 2$,  $1 + \bm{\eta}_{1}^{\prime}\textbf{K}_{1}^{*-1}\bm{\eta}_{1}/2 + \sum_{t=2}^{T}(\bm{\eta}_{t}-\textbf{M}_{t}\bm{\eta}_{t-1})^{\prime}\textbf{W}_{t}^{*-1}(\bm{\eta}_{t}-\textbf{M}_{t}\bm{\eta}_{t-1})/2$), and IG($n/2 + 2$,  $1 + \bm{\xi}_{t}^{\prime}\bm{\xi}_{t}/2$) (for $t = 1,...,T$), respectively. 

\section*{Appendix C: List of Industries used in Section 4.3} We list the different industries that were jointly analyzed in Section 4.3 below.
\singlespacing
\begin{enumerate}
\item Agriculture, Forestry, and Fishing and Hunting
\item Mining, Quarrying, and Oil and Gas Extraction
\item Utilities
\item Construction
\item Manufacturing
\item Wholesale Trade
\item Retail Trade
\item Transportation and Warehousing
\item Information
\item Finance and Insurance
\item Real Estate, and Rental and Leasing
\item Professional, Scientific, and Technical Services
\item Management of Companies and Enterprises
\item Administrative, and Support, Waste Management, and Remediation Services
\item Educational Services
\item Health Care and Social Assistance
\item Arts, Entertainment, and Recreation
\item Accommodation and Food Services
\item Public Administration
\item Other Services
\end{enumerate}

\singlespacing
\bibliographystyle{jasa}  
\bibliography{myref}
      
          \newpage
            \begin{figure}[T]
            \begin{center}
            \begin{tabular}{cc}
            \includegraphics[width=8.5cm,height=6cm]{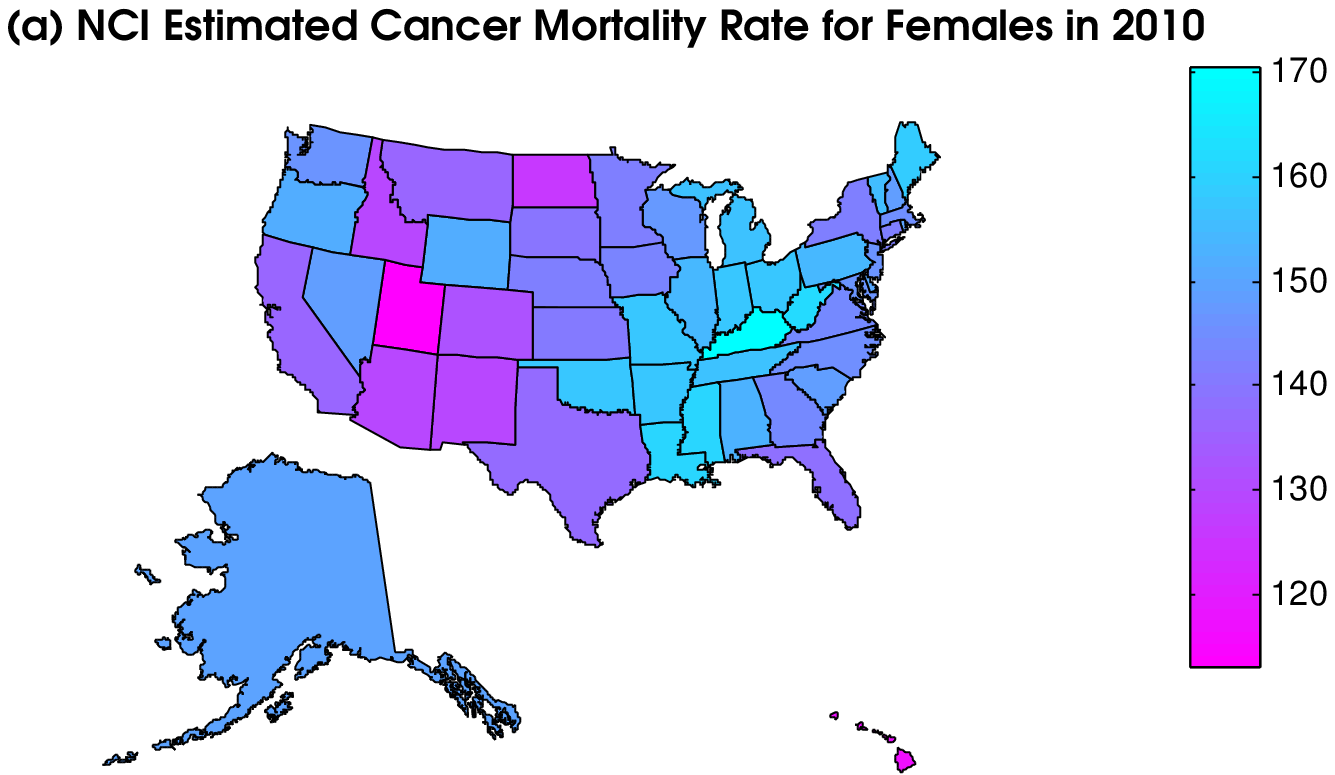}& \includegraphics[width=8.5cm,height=6cm]{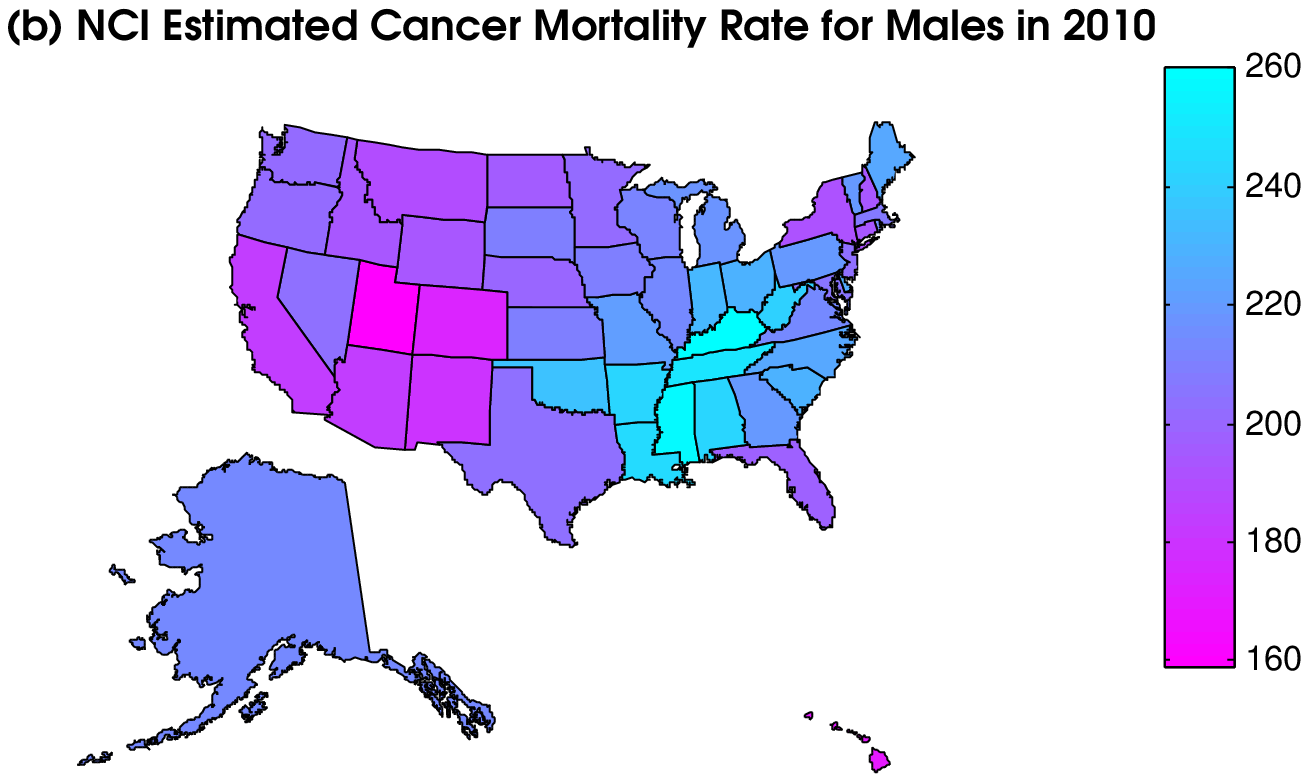}\\
              \includegraphics[width=8.5cm,height=6cm]{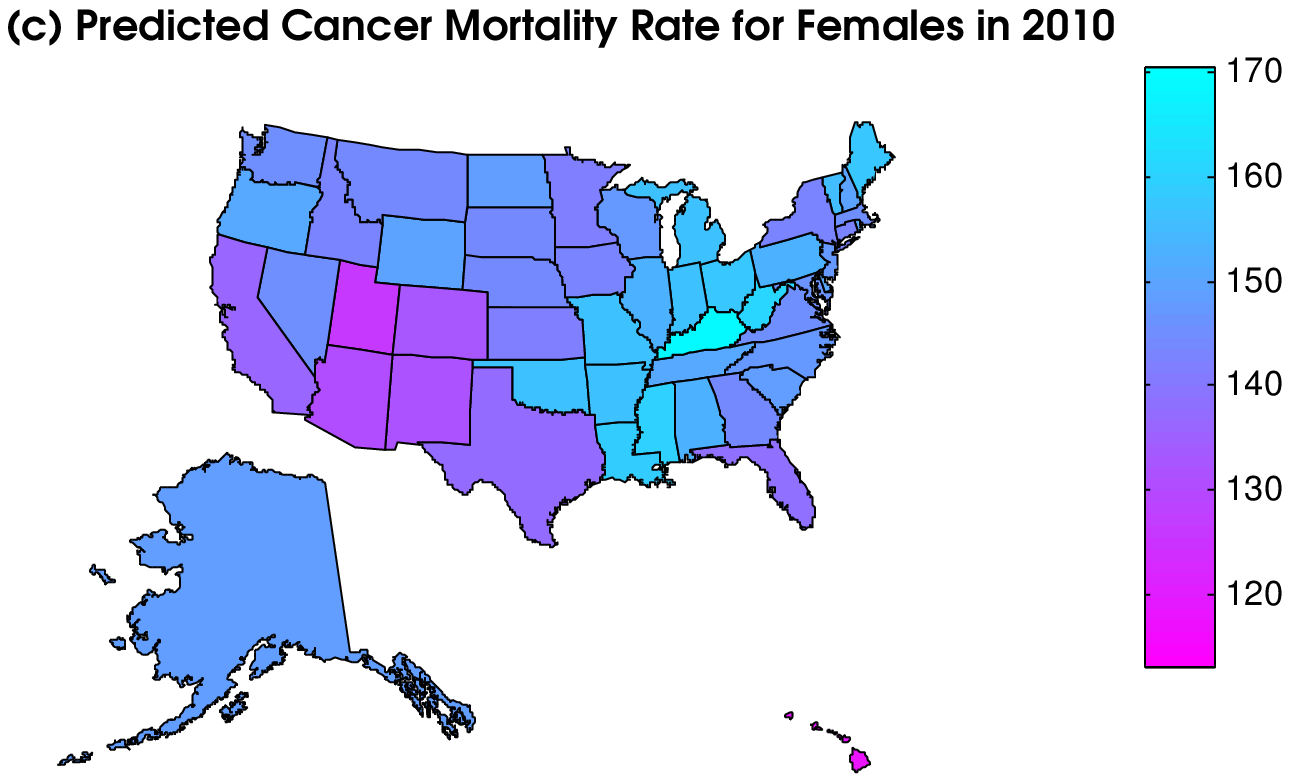}& \includegraphics[width=8.5cm,height=6cm]{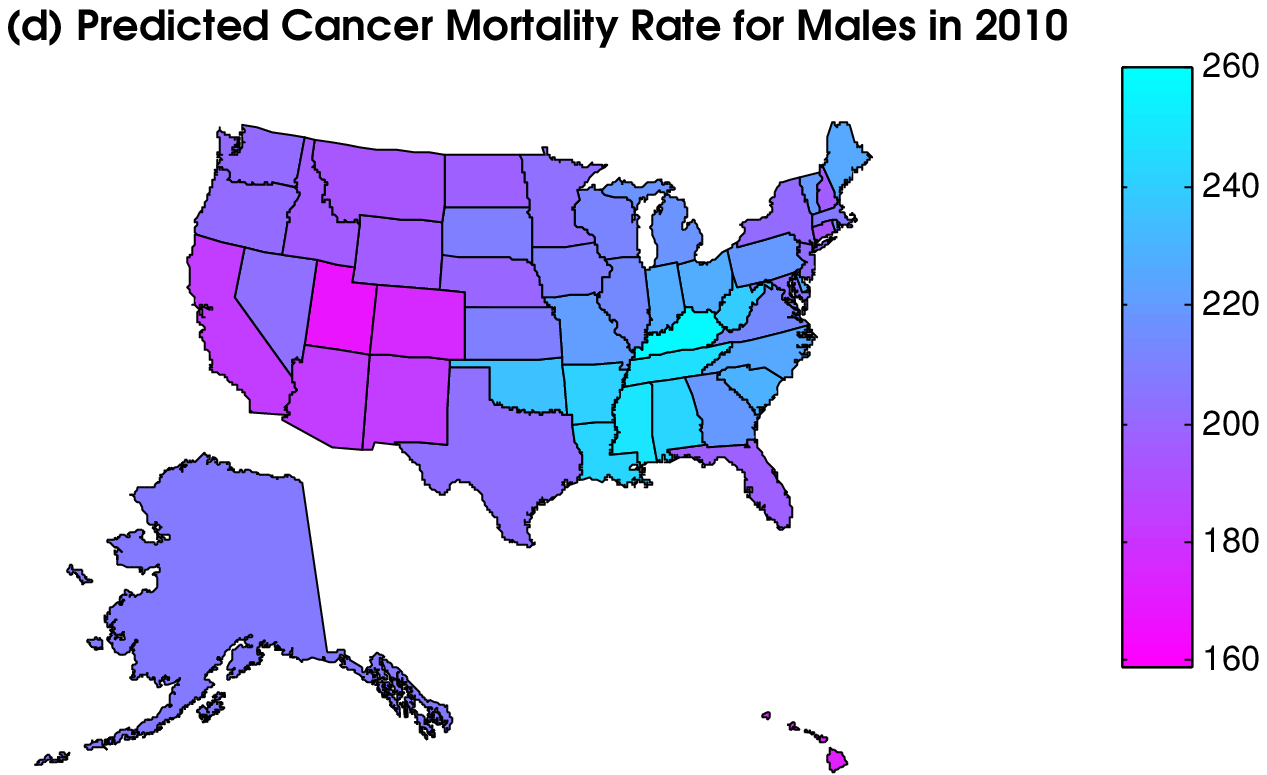}\\ \includegraphics[width=8.5cm,height=6cm]{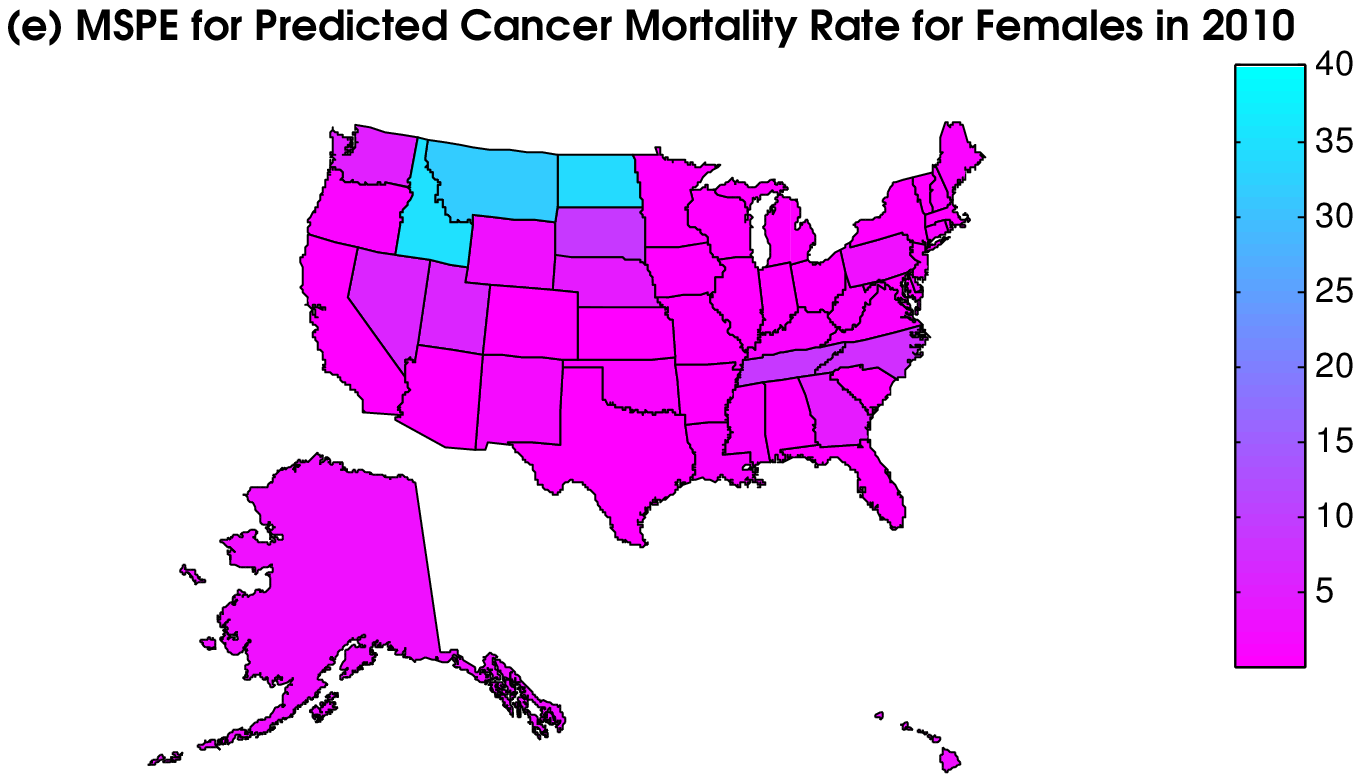}&  \includegraphics[width=8.5cm,height=6cm]{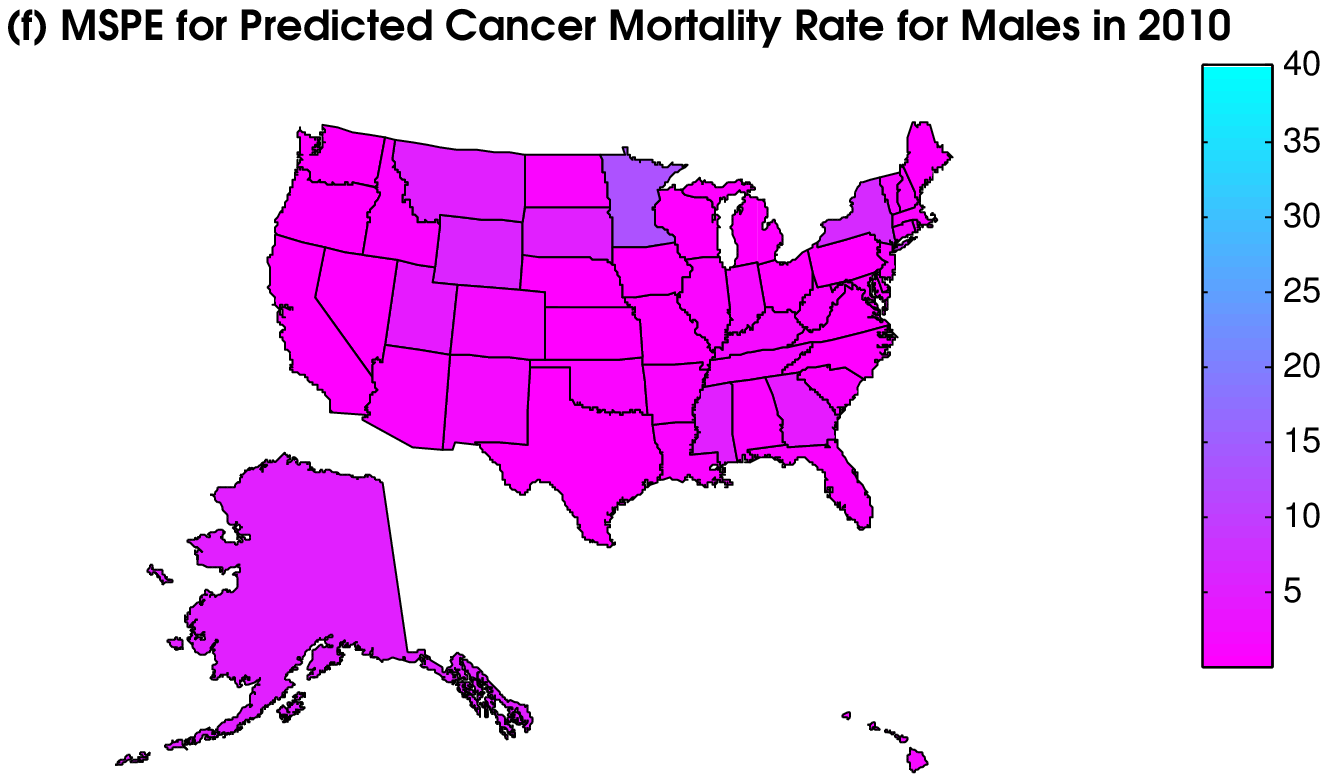}
            \end{tabular}
            \caption{In (a) and (b), we present a selected map of NCI estimates of cancer mortality rates (per 100,000) for women and men, respectively. These values are recorded over each state in the US in 2010. In the second and third rows we present selected posterior means and MSPE of $Y_{t}^{(1)}(\cdot)$ (i.e., mortality rates for women per 100,000) and $Y_{t}^{(2)}(\cdot)$ (i.e., mortality rates for men per 100,000) for each state in the US in 2010, respectively. Notice that the color-scales are different for each panel.}
            \end{center}
            \end{figure}

          	\newpage
            \begin{figure}[T]
            \begin{center}
            \begin{tabular}{cc}
            \includegraphics[width=12cm,height=10cm]{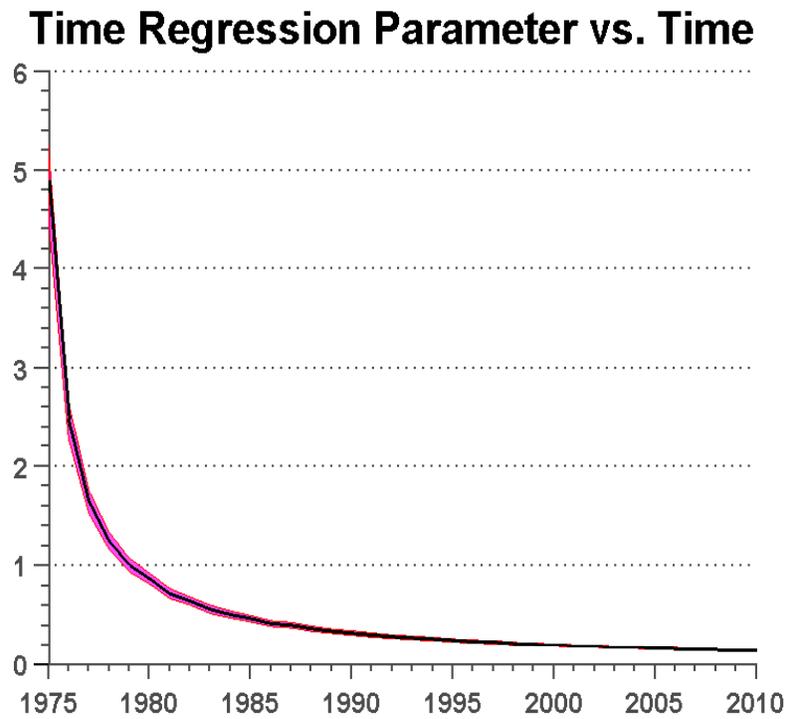}
            \end{tabular}
            \caption{Plot of the posterior mean of $\beta_{3,t}$ versus $t$, where $\bm{\beta}_{t} = (\beta_{1,t},\ldots,\beta_{7,t})^{\prime}$. For the application in Section~\ref{sec:cancer}, $\{\beta_{3,t}\}$ is the regression parameter associated with time. The magenta shaded area indicates the {95$\%$ point-wise (over $\{\beta_{3,t}\}$) credible interval}.}\label{fig2}
            \end{center}
            \end{figure}

\newpage
  \begin{figure}[T]
  \begin{center}
  \begin{tabular}{ccc}
   \includegraphics[width=4.45cm,height=4.45cm]{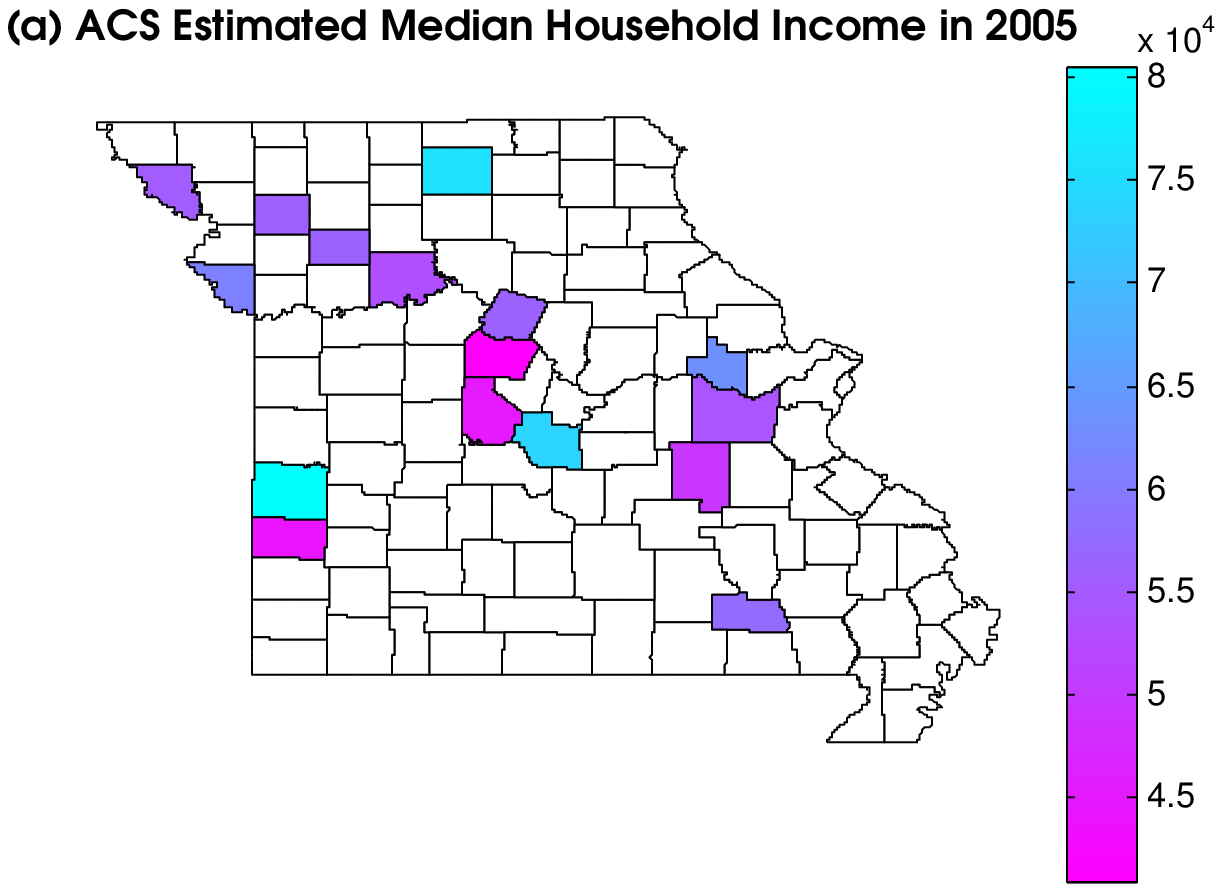}&   \includegraphics[width=4.45cm,height=4.45cm]{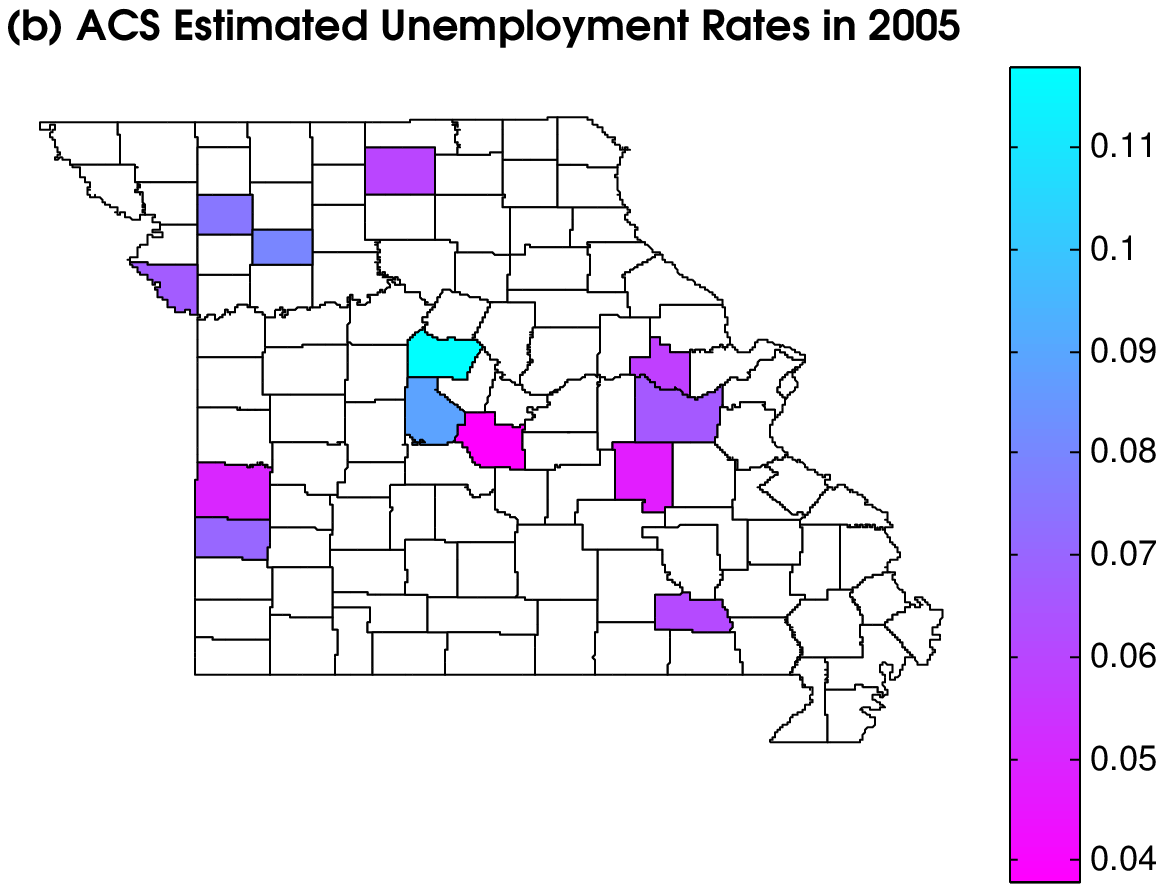}&  \includegraphics[width=4.45cm,height=4.45cm]{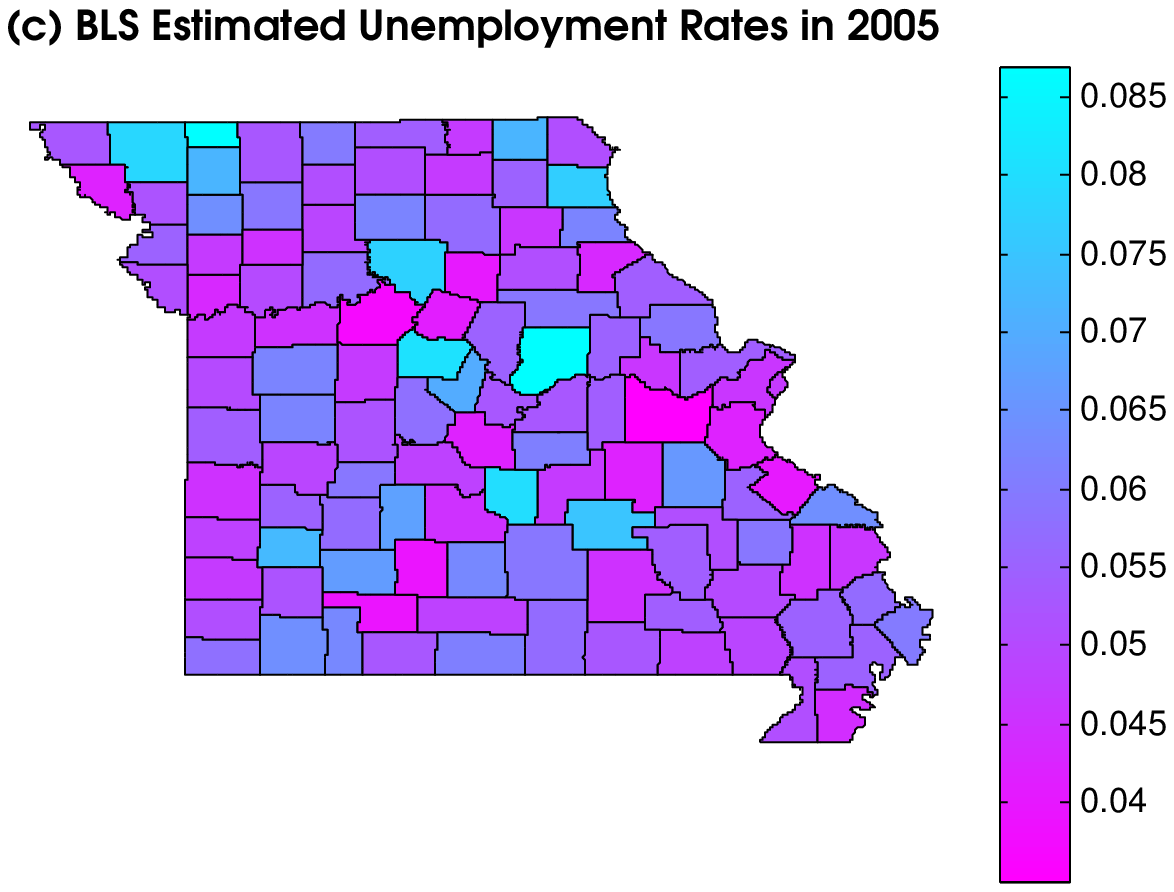}\\
   \includegraphics[width=4.45cm,height=4.45cm]{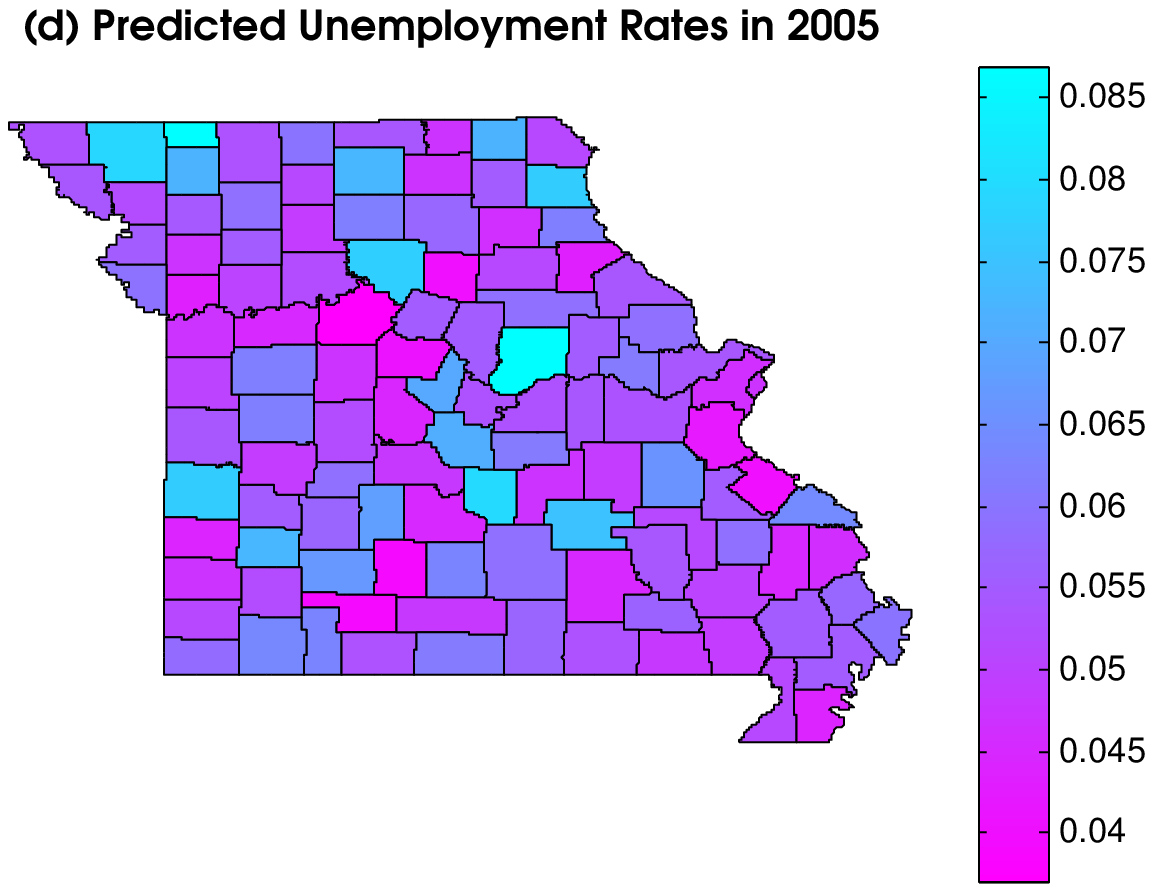}& \includegraphics[width=4.45cm,height=4.45cm]{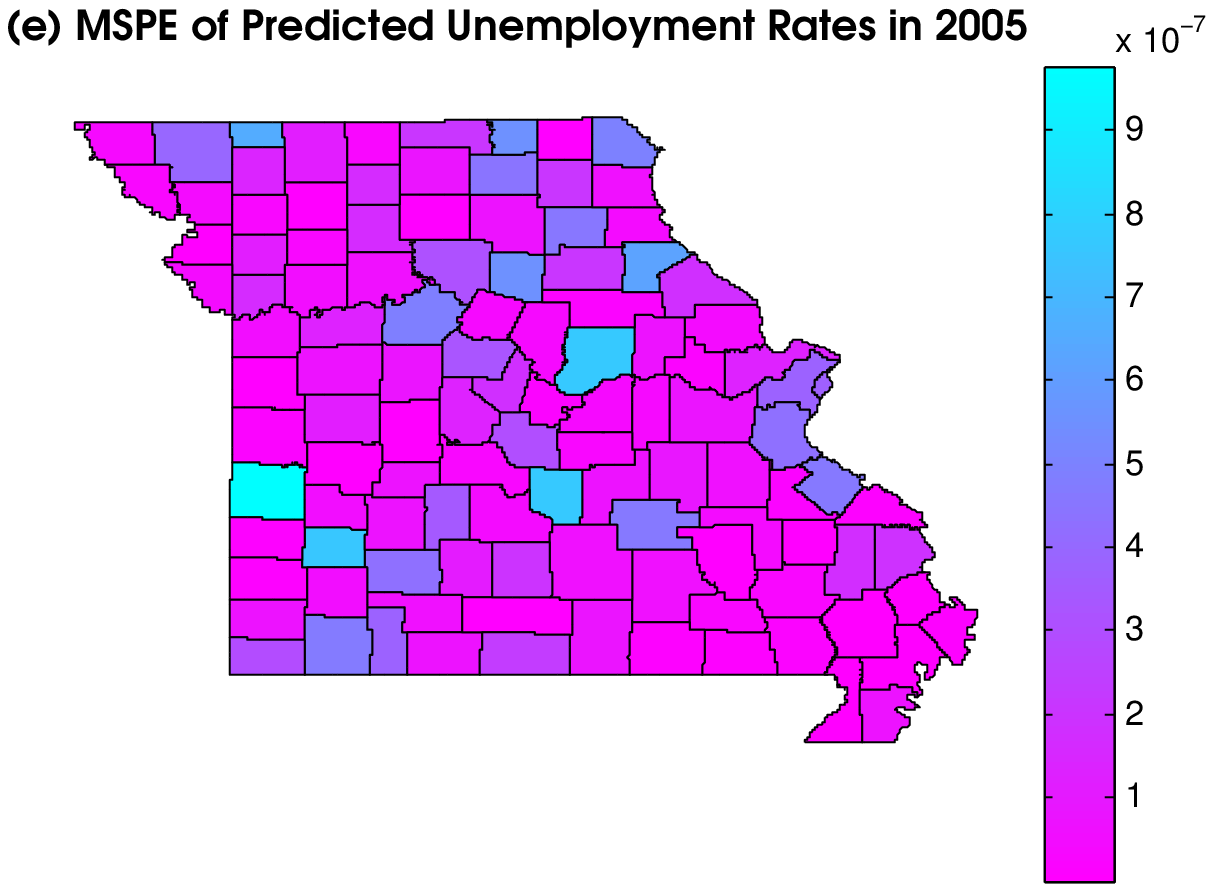}& \hfill
  \end{tabular}
  \caption{In (a) and (b), we present selected maps of ACS 1$\--$year period estimates of the median household income and unemployment rates recorded over Missouri counties in 2005. In (c), we present a selected map of LAUS estimates of the unemployment rates recorded over Missouri counties in 2005. ACS does not provide 1$\--$year period estimates at every county in Missouri; these counties are shaded white. In (d) and (e) we present a selected map of the posterior mean of $Y_{27}^{(1)}(\cdot)$ (i.e., unemployment rates), and the respective posterior MSPE. The color-scales are different for each panel.}
  \end{center}
  \end{figure}

\newpage
  \begin{figure}[T]
  \begin{center}
  \begin{tabular}{c}
   \includegraphics[width=12cm,height=12cm]{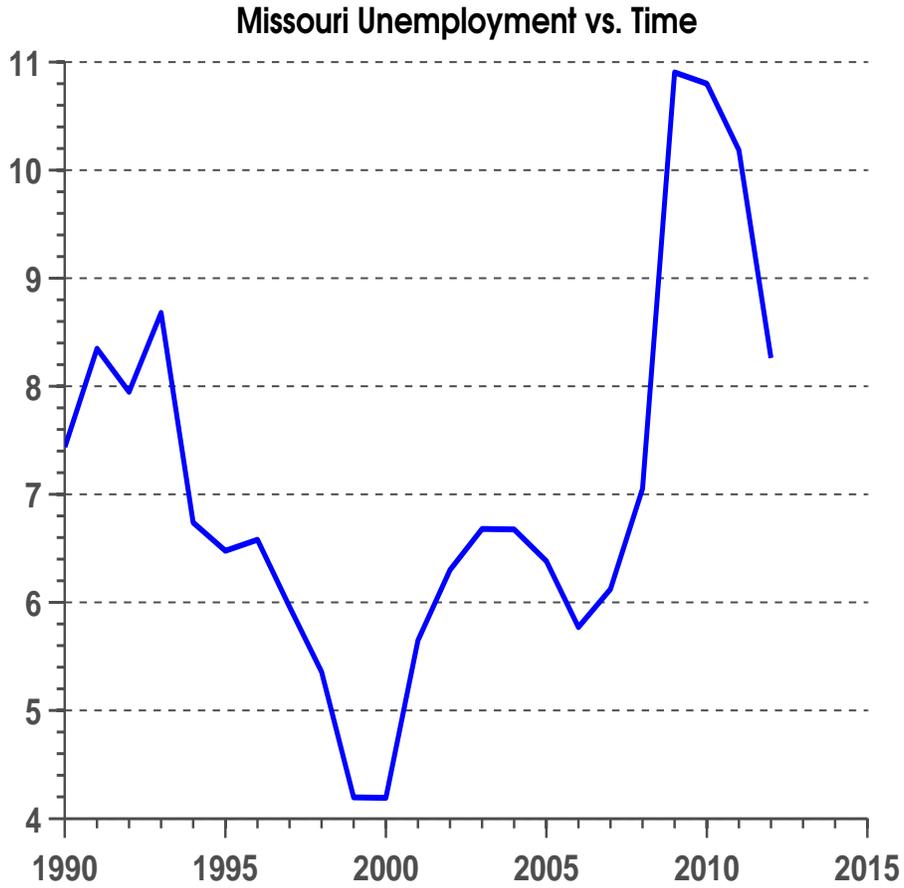}
  \end{tabular}
  \caption{Average predicted Missouri unemployment percentage versus time. Here, by average predicted Missouri unemployment percentage versus time we mean $100\times (1/|D_{\mathrm{P},t}^{(1)}|)\times\sum_{A}\widehat{Y}_{t}^{(1)}(A)$, where $\widehat{Y}_{t}^{(1)}(A)$ is the predicted unemployment rate at year $t$ and county $A$.}
  \end{center}
  \end{figure}

  \newpage
    \begin{figure}[T]
    \begin{center}
    \begin{tabular}{cc}
       \includegraphics[width=8.5cm,height=6cm]{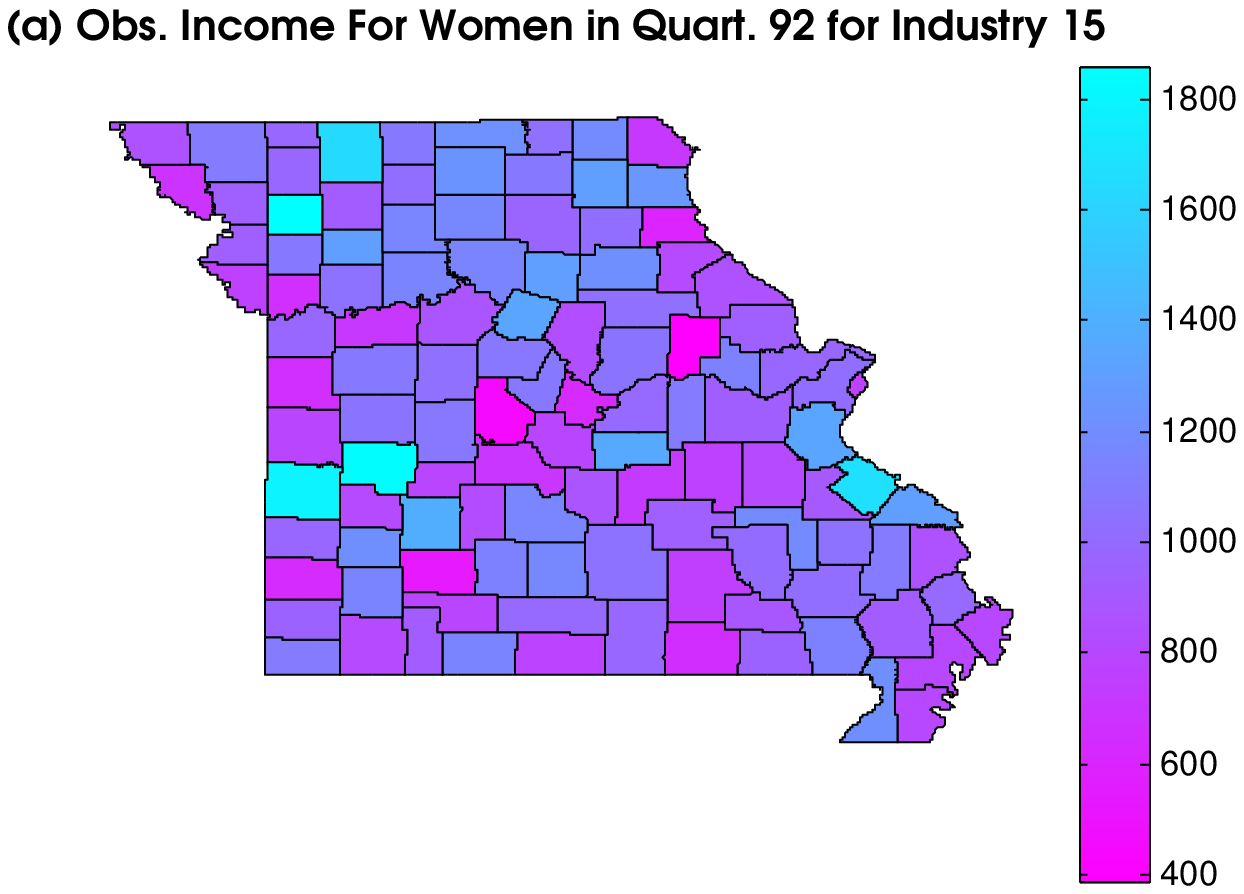}&
       \includegraphics[width=8.5cm,height=6cm]{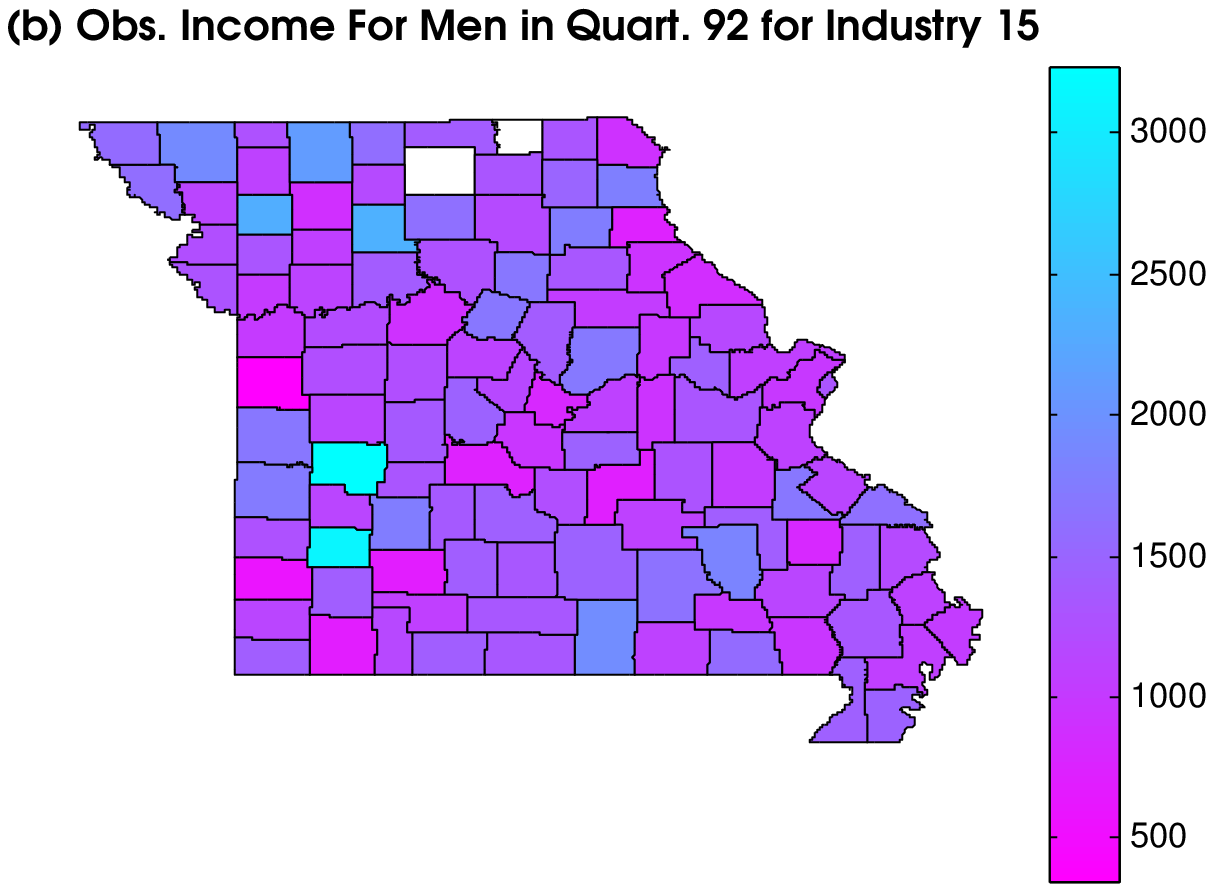}\\
 \includegraphics[width=8.5cm,height=6cm]{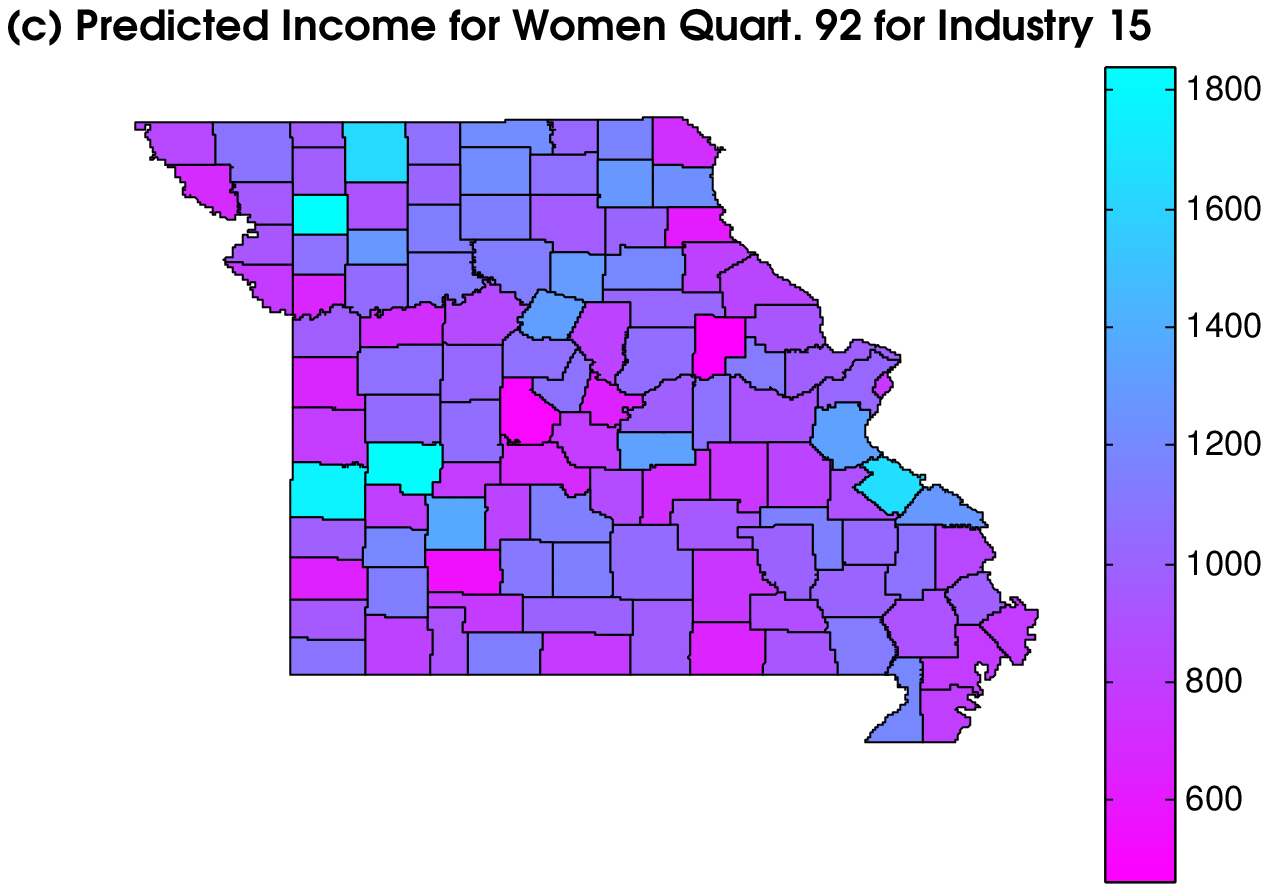}& \includegraphics[width=8.5cm,height=6cm]{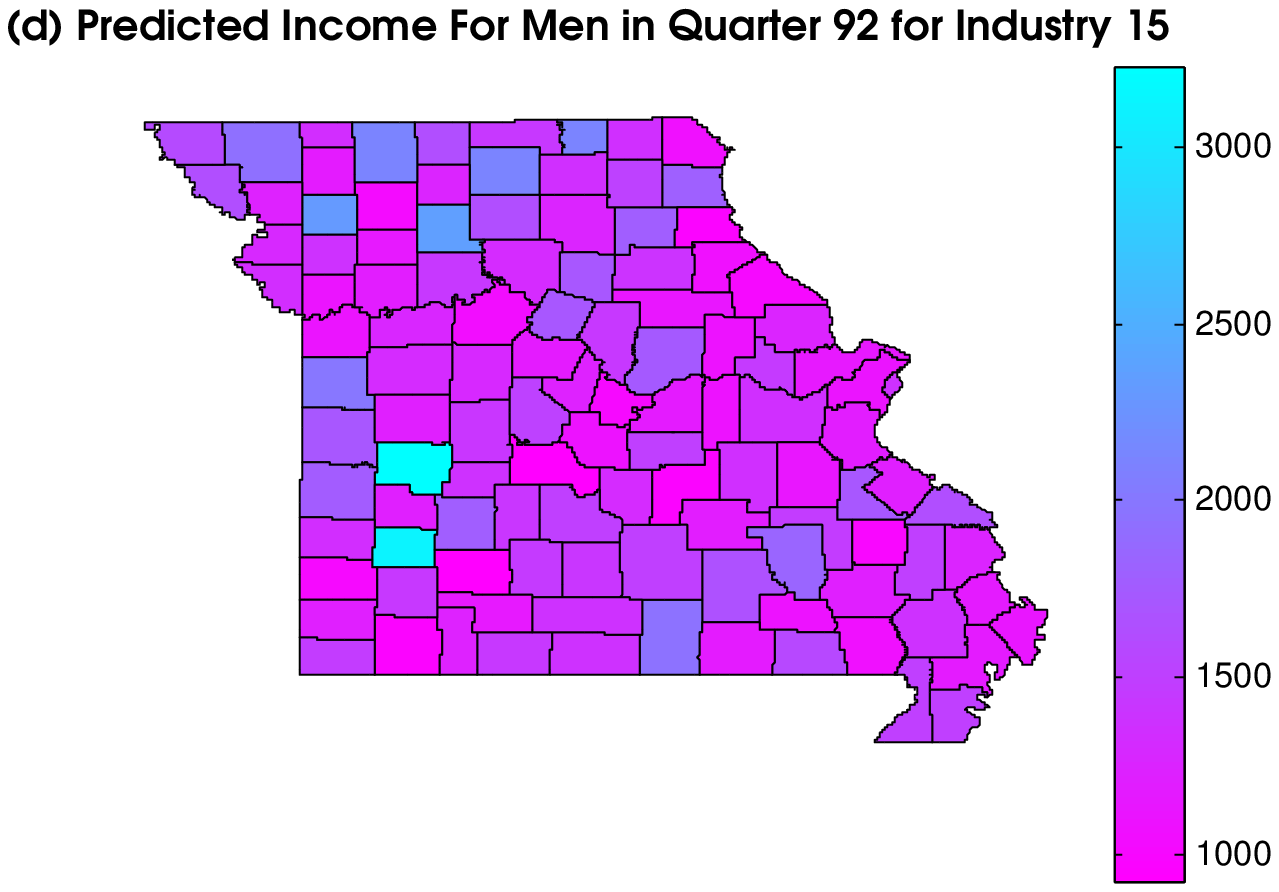}\\
  \includegraphics[width=8.5cm,height=6cm]{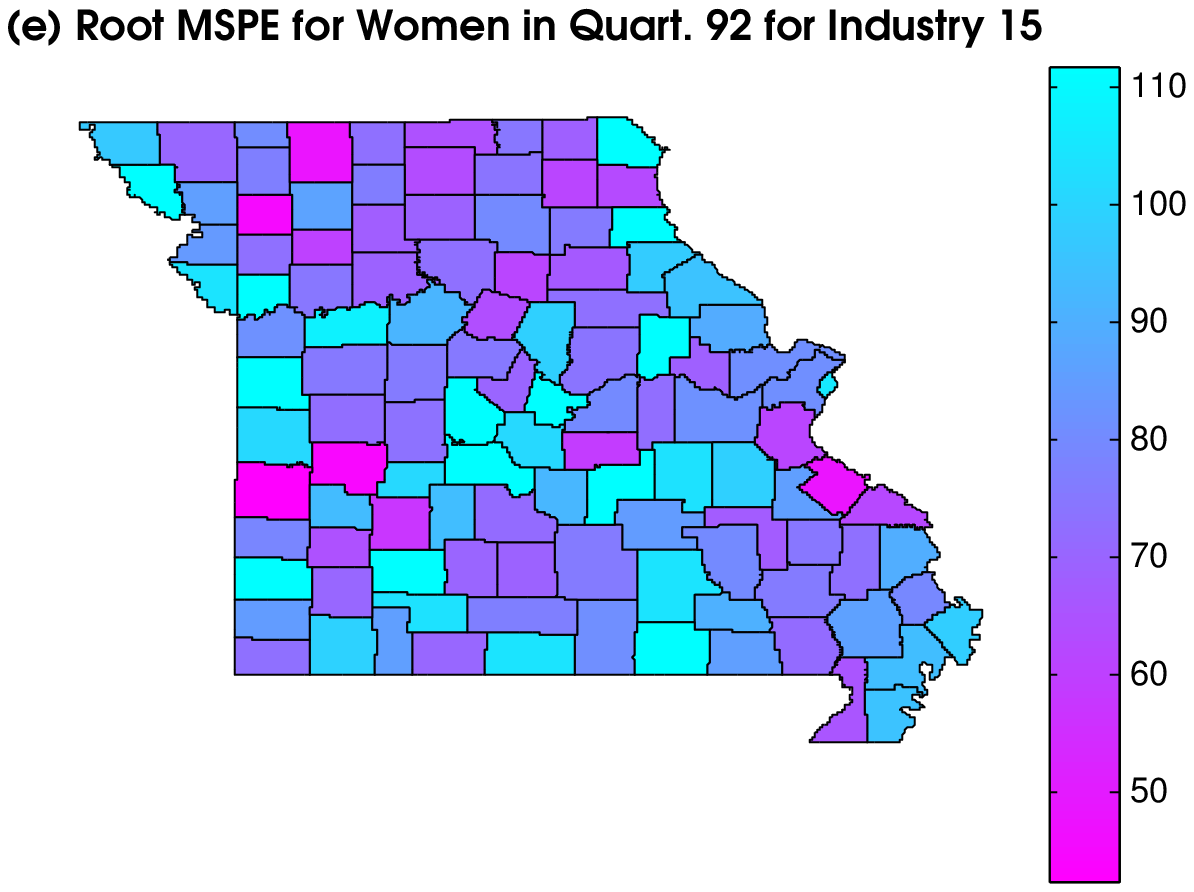}&  \includegraphics[width=8.5cm,height=6cm]{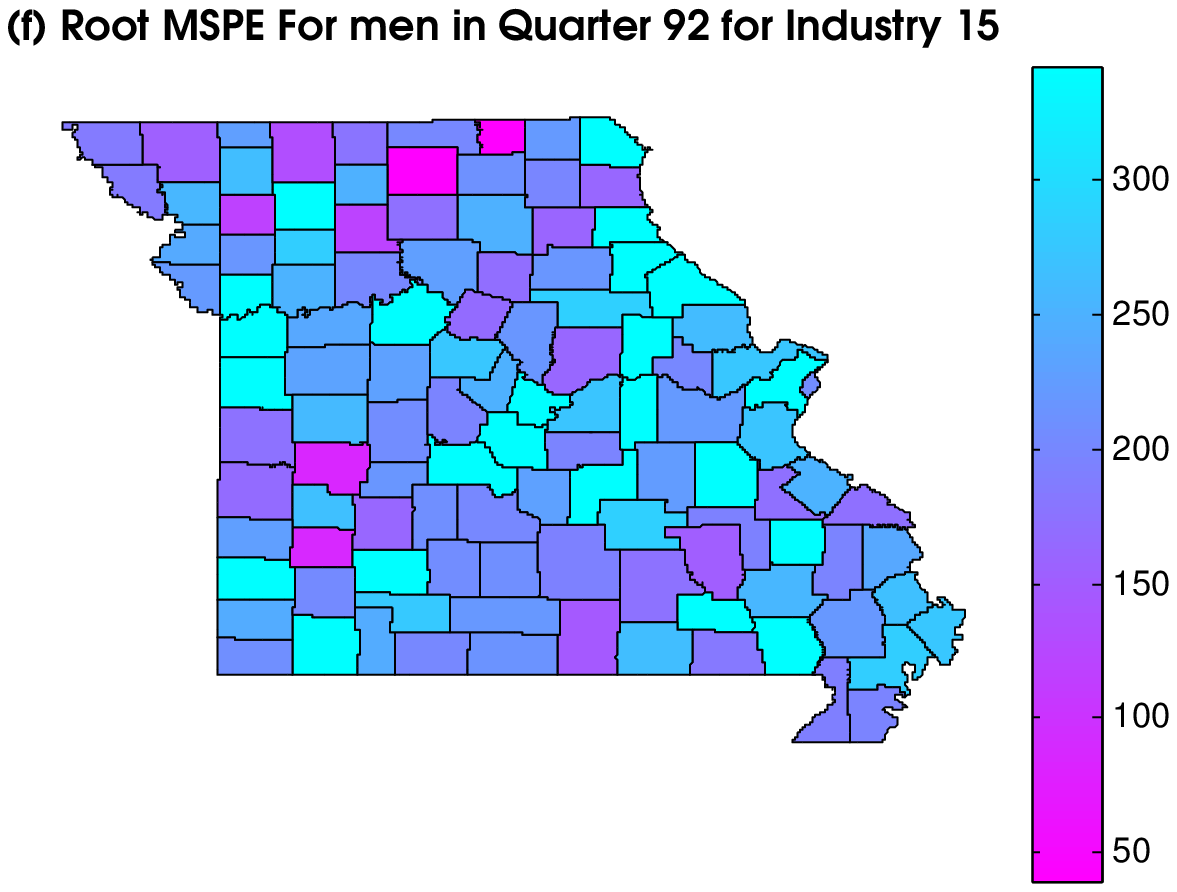}
    \end{tabular}
    \caption{In (a) and (b), we present LEHD estimated average monthly income ({US dollars}) for the state of Missouri, for each gender, for the education industry, and for quarter 92. LEHD does not provide estimates at every county in the US at every quarter; these counties are shaded white. In (c)$\--$(f), we present the corresponding maps (for the state of Missouri, for each gender, for the education industry, and for quarter 92) of predicted monthly income ({US dollars}), and their respective posterior square root MSPE. Notice that the color-scales are different for each panel.}
    \end{center}
    \end{figure}
    
 \end{document}

%% file: notations4.tex
\def\bu{\textbf{u}}

\def\bz{\textbf{z}}

\def\by{\textbf{y}}

\def\bfbeta{\bm{\beta}}

%% file: MSTM_maindoc_and_supp.bbl
\begin{thebibliography}{59}
\newcommand{\enquote}[1]{``#1''}
\expandafter\ifx\csname natexlab\endcsname\relax\def\natexlab#1{#1}\fi

\bibitem[\protect\citename{Banerjee et~al., }2004]{banerjee-etal-2004}
Banerjee, S., Carlin, B.~P., and Gelfand, A.~E. (2004).
\newblock {\em Hierarchical Modeling and Analysis for Spatial Data\/}.
\newblock London, UK: Chapman and Hall.

\bibitem[\protect\citename{Banerjee et~al., }2008]{banerjee}
Banerjee, S., Gelfand, A.~E., Finley, A.~O., and Sang, H. (2008).
\newblock \enquote{Gaussian predictive process models for large spatial data
  sets.}
\newblock {\em Journal of the Royal Statistical Society, Series B\/}, 70,
  825--848.

\bibitem[\protect\citename{Bell and Hillmer, }1990]{Bell}
Bell, W. and Hillmer, S. (1990).
\newblock \enquote{The time series approach to estimation for repeated
  surveys.}
\newblock {\em Survey Methodology\/}, 16, 195--215.

\bibitem[\protect\citename{Berliner, }1996]{berlinhier}
Berliner, L.~M. (1996).
\newblock {\em Hierarchical Bayesian Time-Series Models\/}.
\newblock Kluwer Academic Publishers, Dordrecht, NL.

\bibitem[\protect\citename{Bryant and Graham, }2013]{Bryant}
Bryant, J.~R. and Graham, P.~J. (2013).
\newblock \enquote{Bayesian demographic accounts: Subnational population
  estimation using multiple data Sources.}
\newblock {\em Bayesian Analysis\/}, 8, 1--34.

\bibitem[\protect\citename{Carlin and Banerjee, }2002]{carlinmst}
Carlin, B. and Banerjee, S. (2002).
\newblock \enquote{Hierarchical multivariate CAR models for spatio-temporally
  correlated survival data (with discussion).}
\newblock {\em Bayesian Statistics\/}, 7, 45–63.

\bibitem[\protect\citename{Carter and Kohn, }1994]{cart1994}
Carter, C. and Kohn, R. (1994).
\newblock \enquote{On Gibbs sampling for state space models.}
\newblock {\em Biometrika\/}, 81, 541--553.

\bibitem[\protect\citename{Chen et~al., }2012]{cancertime}
Chen, H., Portier, K., Ghosh, K., Naishadham, D., Kim, H., Zhu, L., Pickle, L.,
  Krapcho, M., Scoppa, S., Jemal, A., and Feuer, E. (2012).
\newblock \enquote{Predicting US and state-level cancer counts for the current
  calendar year.}
\newblock {\em Cancer\/}, 118, 1091–1099.

\bibitem[\protect\citename{Congdon, }2002]{congdon}
Congdon, P. (2002).
\newblock \enquote{A Multivariate Model for Spatio-temporal Health Outcomes
  with an Application to Suicide Mortality.}
\newblock {\em Geographical Analysis\/}, 36, 235--258.

\bibitem[\protect\citename{Cressie, }1993]{cressie}
Cressie, N. (1993).
\newblock {\em Statistics for Spatial Data, \textup{rev. edn}\/}.
\newblock New York, NY: Wiley.

\bibitem[\protect\citename{Cressie and Huang, }1999]{HuangCressie2}
Cressie, N. and Huang, H. (1999).
\newblock \enquote{Classes of nonseparable, spatio-temporal stationary
  covariance functions.}
\newblock {\em Journal of the American Statistical Association\/}, 94,
  1330--1340.

\bibitem[\protect\citename{Cressie and Johannesson, }2008]{johan}
Cressie, N. and Johannesson, G. (2008).
\newblock \enquote{Fixed rank kriging for very large spatial data sets.}
\newblock {\em Journal of the Royal Statistical Society, Series B\/}, 70,
  209--226.

\bibitem[\protect\citename{Cressie et~al., }2010]{kang-cressie-shi-2010}
Cressie, N., Shi, T., and Kang, E.~L. (2010).
\newblock \enquote{{ Using temporal variability to improve spatial mapping with
  application to satellite data}.}
\newblock {\em Canadian Journal of Statistics\/}, 38, 271--289.

\bibitem[\protect\citename{Cressie and Wikle, }2011]{cressie-wikle-book}
Cressie, N. and Wikle, C.~K. (2011).
\newblock {\em Statistics for Spatio-Temporal Data\/}.
\newblock Hoboken, NJ: Wiley.

\bibitem[\protect\citename{Dahlhaus, }1997]{Dahlhaus}
Dahlhaus, R. (1997).
\newblock \enquote{Fitting time series models to nonstationary processes.}
\newblock {\em The Annals of Statistics\/}, 25, 1--37.

\bibitem[\protect\citename{Daniels et~al., }2006]{daniels}
Daniels, M., Zhou, Z., and Zou, H. (2006).
\newblock \enquote{Conditionally specified space–time models for multivariate
  processes.}
\newblock {\em Journal of Computational and Graphical Statistics\/}, 15,
  157--177.

\bibitem[\protect\citename{Elliott and Davis, }2013]{Elliott}
Elliott, M. and Davis, W. (2013).
\newblock \enquote{Obtaining cancer risk factor prevalence estimates in small
  areas: Combining data from two surveys.}
\newblock {\em Journal of the Royal Statistical Society, Series C\/}, 54,
  595--609.

\bibitem[\protect\citename{Etxeberria1 et~al., }2014]{cancertime2}
Etxeberria1, J., Goicoa1, T., Ugarte1, M., and Militino, A. (2014).
\newblock \enquote{Evaluating space-time models for short-term cancer mortality
  risk predictions in small areas.}
\newblock {\em Biometrical Journal\/}, 56, 383--402.

\bibitem[\protect\citename{Feder, }2013]{Feder}
Feder, M. (2013).
\newblock \enquote{Time series analysis of repeated surveys: The state-space
  approach.}
\newblock {\em Statistica Neerlandica\/}, 55, 182--199.

\bibitem[\protect\citename{Finley et~al., }2010]{finley2}
Finley, A.~O., Banerjee, S., Waldmann, P., and Ericsson, T. (2010).
\newblock \enquote{Hierarchical spatial process models for multiple traits in
  large genetic trials.}
\newblock {\em Journal of the American Statistical Association\/}, 105,
  506--521.

\bibitem[\protect\citename{Finley et~al., }2009]{finley}
Finley, A.~O., Sang, H., Banerjee, S., and Gelfand, A.~E. (2009).
\newblock \enquote{Improving the performance of predictive process modeling for
  large datasets.}
\newblock {\em Computational Statistics and Data Analysis\/}, 53, 2873--2884.

\bibitem[\protect\citename{Fr{\"u}wirth-Schnatter, }1994]{schnatter94}
Fr{\"u}wirth-Schnatter, S. (1994).
\newblock \enquote{Data augmentation and dynamic linear models.}
\newblock {\em Journal of Time Series Analysis\/}, 15, 183--202.

\bibitem[\protect\citename{Giorgi et~al., }2013]{Giorgi}
Giorgi, E., Sesay, S., Terlouw, D., and Diggle, P. (2013).
\newblock \enquote{Combining data from multiple spatially referenced prevalence
  surveys using generalized linear geostatistical models.}
\newblock {\em arXiv preprint arXiv: 1308.2790\/}.

\bibitem[\protect\citename{Gneiting, }1992]{Gneitingcorr}
Gneiting, T. (1992).
\newblock \enquote{Correlation functions for atmospheric data analysis.}
\newblock {\em Quarterly Journal of the Royal Meteorological Society\/}, 125,
  2449--2464.

\bibitem[\protect\citename{Griffith, }2000]{griffith2000}
Griffith, D. (2000).
\newblock \enquote{A linear regression solution to the spatial autocorrelation
  problem.}
\newblock {\em Journal of Geographical Systems\/}, 2, 141--156.

\bibitem[\protect\citename{Griffith, }2002]{griffith2002}
--- (2002).
\newblock \enquote{A spatial filtering specification for the auto-Poisson
  model.}
\newblock {\em Statistics and Probability Letters\/}, 58, 245--251.

\bibitem[\protect\citename{Griffith, }2004]{griffith2004}
--- (2004).
\newblock \enquote{A spatial filtering specification for the auto-logistic
  model.}
\newblock {\em Environment and Planning A\/}, 36, 1791--1811.

\bibitem[\protect\citename{Griffith and Tiefelsdorf, }2007]{griffith2007}
Griffith, D. and Tiefelsdorf, M. (2007).
\newblock \enquote{Semiparametric filtering of spatial autocorrelation: The
  eigenvector approach.}
\newblock {\em Environment and Planning A\/}, 39, 1193--1221.

\bibitem[\protect\citename{Higham, }1988]{Higham}
Higham, N. (1988).
\newblock \enquote{Computing a nearest symmetric positive semidefinite matrix.}
\newblock {\em Linear Algebra and its Applications\/}, 105, 103--118.

\bibitem[\protect\citename{Holt, }2009]{holt}
Holt, J. (2009).
\newblock \enquote{A Summary of the Primary Causes of the Housing Bubble and
  the Resulting Credit Crisis: A Non-Technical Paper.}
\newblock {\em The Journal of Business Inquiry\/}, 8, 120--129.

\bibitem[\protect\citename{Hughes and Haran, }2013]{hughes}
Hughes, J. and Haran, M. (2013).
\newblock \enquote{Dimension reduction and alleviation of confounding for
  spatial generalized linear mixed model.}
\newblock {\em Journal of the Royal Statistical Society, Series B\/}, 75,
  139--159.

\bibitem[\protect\citename{Jones, }2010]{Jones}
Jones, R. (2010).
\newblock \enquote{Best linear unbiased estimators for repeated surveys.}
\newblock {\em Journal of the Royal Statistical Society, Series B\/}, 42,
  221--226.

\bibitem[\protect\citename{Keller and Olkin, }2002]{Keller}
Keller, T. and Olkin, I. (2002).
\newblock {\em Combining correlated unbiased estimators of the mean of a normal
  distribution\/}.
\newblock Tech. Report, National Agricultural Statistics Service.

\bibitem[\protect\citename{Kern and Borgman, }2008]{dorfman}
Kern, J. and Borgman, L. (2008).
\newblock {\em The two sample problem\/}.
\newblock Tech. Report, US Bureau of Labor Statistics.

\bibitem[\protect\citename{Lindgren et~al., }2011]{lindgren-2011}
Lindgren, F., Rue, H., and Lindstr\"{o}m, J. (2011).
\newblock \enquote{An explicit link between Gaussian fields and Gaussian Markov
  random fields: The stochastic partial differential equation approach.}
\newblock {\em Journal of the Royal Statistical Society, Series B\/}, 73,
  423--498.

\bibitem[\protect\citename{Lohr and Brick, }2012]{lohr}
Lohr, S. and Brick, M. (2012).
\newblock \enquote{Blending domain estimates from two victimization surveys
  with possible bias.}
\newblock {\em The Canadian Journal of Statistics\/}, 40, 679--969.

\bibitem[\protect\citename{Merkouris, }2012]{Merkouris1}
Merkouris, T. (2012).
\newblock \enquote{Combining independent regression estimators from multiple
  surveys.}
\newblock {\em Journal of the American Statistical Association\/}, 99,
  1131--1139.

\bibitem[\protect\citename{Nychka et~al., }2014]{nychkaLK}
Nychka, D., Bandyopadhyay, S., Hammerling, D., Lindgren, F., and Sain, S.
  (2014).
\newblock \enquote{A Multi-resolution Gaussian process model for the analysis
  of large spatial data sets.}
\newblock {\em Journal of Computational and Graphical Statistics\/},  DOI:
  10.1080/10618600.2014.914946.

\bibitem[\protect\citename{Oehlert, }1992]{delta}
Oehlert, G. (1992).
\newblock \enquote{A note on the delta method.}
\newblock {\em The American Statistician\/}, 46, 27--29.

\bibitem[\protect\citename{Pettitt et~al., }2002]{pettitt}
Pettitt, A., Weir, I., and Hart, A. (2002).
\newblock \enquote{A conditional autoregressive Gaussian process for
  irregularly spaced multivariate data with application to modelling large sets
  of binary data.}
\newblock {\em Statistics and Computing\/}, 12, 353--367.

\bibitem[\protect\citename{Porter et~al., }2013]{aaronp}
Porter, A., Holan, S.~H., and Wikle, C.~K. (2013).
\newblock \enquote{Small area estimation via multivariate Fay-Herriot models
  with latent spatial dependence.}
\newblock {\em arXiv preprint arXiv: 1310.7211\/}.

\bibitem[\protect\citename{Porter et~al., }2014]{aaronpBayes}
--- (2014).
\newblock \enquote{Bayesian Semiparametric Hierarchical Empirical Likelihood
  Spatial Models.}
\newblock {\em arXiv preprint arXiv: 1405.3880\/}.

\bibitem[\protect\citename{Raghunathan et~al., }2007]{Raghunathan}
Raghunathan, T., Xie, D., Schenker, N., Parsons, V., Davis, W., Dodd, K., and
  Feuer, E. (2007).
\newblock \enquote{Combining information from two surveys to estimate
  county-level prevalence rates of cancer risk factors and screening.}
\newblock {\em Journal of the American Statistical Association\/}, 102,
  1131--1139.

\bibitem[\protect\citename{Ravishanker and Dey, }2002]{ravishank}
Ravishanker, N. and Dey, D.~K. (2002).
\newblock {\em A First Course in Linear Model Theory\/}.
\newblock Boca Raton, FL: Chapman and Hall/CRC.

\bibitem[\protect\citename{Reich et~al., }2006]{Reich}
Reich, B., Hodges, J., and Zadnik, V. (2006).
\newblock \enquote{Effects of residual smoothing on the posterior of the fixed
  effects in disease-mapping models.}
\newblock {\em Biometrics\/}, 62, 1197--1206.

\bibitem[\protect\citename{Royle and Berliner, }1999]{berlinerroyle}
Royle, A. and Berliner, M. (1999).
\newblock \enquote{A hierarchical approach to multivariate spatial modeling and
  prediction.}
\newblock {\em Journal of Agricultural, Biological, and Environmental
  Statistics\/}, 19, 2.

\bibitem[\protect\citename{Royle et~al., }1999]{royle1999}
Royle, J., Berliner, M., Wikle, C., and Milliff, R. (1999).
\newblock \enquote{A hierarchical spatial model for constructing wind fields
  from scatterometer data in the Labrador sea.}
\newblock In {\em Case Studies in Bayesian Statistics\/}, eds. C.~Gatsonis,
  R.~Kass, B.~Carlin, A.~Carriquiry, A.~Gelman, I.~Verdinelli, and M.~West,
  367--382. Springer New York.

\bibitem[\protect\citename{Sampson and Guttorp, }1992]{guttorpandpeterson}
Sampson, P. and Guttorp, P. (1992).
\newblock \enquote{Nonparametric estimation of nonstationary spatial covariance
  structure.}
\newblock {\em Journal of the American Statistical Association\/}, 87,
  108--119.

\bibitem[\protect\citename{Shumway and Stoffer, }2006]{shumway}
Shumway, R. and Stoffer, D. (2006).
\newblock {\em Time Series Analysis and Its Applications: With R Examples\/}.
\newblock New York, NY, USA: Springer.

\bibitem[\protect\citename{Speilman et~al., }2013]{speilman}
Speilman, S., Folch, D., and Nagle, N. (2013).
\newblock \enquote{Patterns and causes of uncertainty in the American Community
  Survey.}
\newblock {\em Applied Geography\/}, 46, 147--157.

\bibitem[\protect\citename{Stein, }2005]{steinSep}
Stein, M. (2005).
\newblock \enquote{Space-time covariance functions.}
\newblock {\em Journal of the American Statistical Association\/}, 100,
  310--321.

\bibitem[\protect\citename{Stein, }2013]{steinr}
--- (2013).
\newblock \enquote{Limitations on low rank approximations for covariance
  matrices of spatial data.}
\newblock {\em Spatial Statistics\/}, In Press.

\bibitem[\protect\citename{Sun and Li, }2012]{reviewmethods}
Sun, Y. and Li, B. (2012).
\newblock \enquote{Geostatistics for large datasets.}
\newblock In {\em Space-Time Processes and Challenges Related to Environmental
  Problems\/}, eds. E.~Porcu, J.~M. Montero, and M.~Schlather,  55--77.
  Springer.

\bibitem[\protect\citename{Tzala and Best, }2007]{bestmst}
Tzala, E. and Best, N. (2007).
\newblock \enquote{Bayesian latent variable modelling of multivariate
  spatio-temporal variation in cancer mortality.}
\newblock {\em Statistical Methods in Medical Research\/},  1--22.

\bibitem[\protect\citename{Waller et~al., }1997]{stcar}
Waller, L., Carlin, B., Xia, H., and Gelfand, A. (1997).
\newblock \enquote{Hierarchical spatio-temporal mapping of disease rates.}
\newblock {\em Journal of the American Statistical Association\/}, 92,
  607--617.

\bibitem[\protect\citename{Wang et~al., }2012]{wang}
Wang, J., Holan, S., Nandram, B., Barboza, W., Toto, C., and Anderson, E.
  (2012).
\newblock \enquote{A Bayesian approach to estimating agricultural yield based
  on multiple repeated surveys.}
\newblock {\em Journal of Agricultural, Biological, and Environmental
  Statistics\/}, 17, 84--106.

\bibitem[\protect\citename{Wikle et~al., }2001]{wikle2001}
Wikle, C., Milliff, R., Nychka, D., and Berliner, L. (2001).
\newblock \enquote{Spatiotemporal hierarchical Bayesian modeling tropical ocean
  surface winds.}
\newblock {\em Journal of the American Statistical Association (Theory and
  Methods)\/}, 96, 382--397.

\bibitem[\protect\citename{Wikle, }2010]{wikleHandbook}
Wikle, C.~K. (2010).
\newblock \enquote{Low-rank representations for spatial processes.}
\newblock In {\em Handbook of Spatial Statistics\/}, eds. A.~E. Gelfand, P.~J.
  Diggle, M.~Fuentes, and P.~Guttorp,  107--118. Boca Raton, FL: Chapman $\&$
  Hall/CRC Press.

\bibitem[\protect\citename{Zhu et~al., }2002]{zhuglm}
Zhu, J., Eickhoff, C., and Yan, P. (2002).
\newblock \enquote{Generalized Linear Latent Variable Models for Repeated
  Measures of Spatially Correlated Multivariate Data.}
\newblock {\em Biometrics\/}, 61, 674--683.

\end{thebibliography}
